\documentclass[11pt,a4paper]{article}
\usepackage{pub}
\usepackage[T1]{fontenc}
\usepackage[english]{babel}
\usepackage{lmodern}
\usepackage{float}
\usepackage{tikz-feynman}
\usepackage{bbm}
\usepackage{bm}
\usepackage{color}
\usepackage{slashed}
\usepackage{graphicx}
\usepackage{braket}
\usepackage{bbold}
\usepackage{subfigure}
\usepackage{subcaption}
\usepackage{dsfont}
\usepackage{todonotes}
\usepackage[normalem]{ulem}

\DeclareGraphicsExtensions{.pdf,.png,.jpg}

\allowdisplaybreaks[1]
\def\simg{{\ \lower-1.2pt\vbox{\hbox{\rlap{$>$}\lower6pt\vbox{\hbox{$\sim$}}}}\ }}
\def\siml{{\ \lower-1.2pt\vbox{\hbox{\rlap{$<$}\lower6pt\vbox{\hbox{$\sim$}}}}\ }}

\makeatletter \@addtoreset{equation}{section} \makeatother
\renewcommand{\theequation}{\arabic{section}.\arabic{equation}}

\newcommand{\vrel}{v_\rmi{rel}}

\newcommand{\rmi}[1]{{\mbox{\scriptsize #1}}}
\newcommand{\rmii}[1]{{\mbox{\tiny\rm{#1}}}}

\newcommand{\yS}{y_{\scriptscriptstyle S}}
\newcommand{\yV}{y_{\scriptscriptstyle V}}
\newcommand{\DmVS}{\Delta m_{\scriptscriptstyle VS}}

\newcommand{\DmFS}{\Delta m_{\scriptscriptstyle FS}}
\newcommand{\dmFS}{\delta m_{\scriptscriptstyle FS}}
\newcommand{\DmFV}{\Delta m_{\scriptscriptstyle FV}}
\newcommand{\dmVS}{\delta m_{\scriptscriptstyle VS}}
\newcommand{\vE}{v_{\rmii{E}}}

\begin{document}

\flushbottom

\begin{titlepage}
\begin{centering}

\vfill

{\Large{\bf
Lighting up LZ with Loopy Transitions \\
in a leptophilic Dark Sector
}}

\vspace{0.8cm}

S.~Biondini$^a$, C.~Hofheinz$^a$, S.~Melchin$^a$, L.~Tiberi$^{b,c}$ and S.~Vogl$^a$

\vspace{0.8cm}

$^{a}${\em Institute of Physics, University of Freiburg,
\\
Hermann-Herder-Stra{\ss}e 3, 79014 Freiburg, Germany} 
\\
\vspace{0.15 cm}
$^{b}${\em Dipartimento di Fisica e Geologia, Università degli studi di Perugia
Via A. Pascoli, I-06123, Perugia, Italy
}
\\
\vspace{0.15 cm}
$^{c}${\em INFN, Sezione di Perugia,
\\
Via A. Pascoli, I-06123, Perugia, Italy}

\vspace*{0.8cm}

\end{centering}

\vspace*{0.3cm}

\vspace{1cm}
\noindent
\textbf{Abstract}: 
We study the direct detection phenomenology and cosmology of a multicomponent dark sector with a leptophilic t-channel portal. The coupling to right-handed Standard Model leptons is mediated by a charged vector-like fermion, allowing both a scalar and a vector dark candidate to connect to the SM through the same mediator. Motivated by the high-energy nuclear-recoil event recently reported by the LUX-ZEPLIN (LZ) experiment, we consider a nearly degenerate dark sector with mass splittings of order a few hundred~keV and investigate the resulting inelastic transitions. Concretely, we focus on endothermic scattering of the dark states, which can shift the recoil spectrum towards the high-energy region preferred by  LZ while suppressing low-energy events. We scrutinize the effective operators generating the dark-matter--photon coupling, which provides the essential interaction for nuclear recoils. Interestingly, the inelastic transition arises at dimension five whereas elastic scattering on nuclei is induced only at dimension six. This naturally explains the prominence of endothermic scattering compared to its elastic counterpart. We  determine the parameter space compatible with the observed dark matter abundance and scrutinize the  event rate and signal shape for scattering on xenon.  Our results demonstrate that the observation of a single $E_R=248\pm 46$ keV nuclear recoil event in LZ can be accommodated by thermally produced dark matter in a leptophilic dark sector.

\vfill
\end{titlepage}

\clearpage
\pagenumbering{gobble} 
\tableofcontents

\clearpage
\pagenumbering{arabic}
\setcounter{page}{1}

\section{Introduction}
\label{sec:intro}

The existence of dark matter (DM) is firmly established by a wide range of cosmological and astrophysical observations, while its microscopic nature remains unknown. Among the many possibilities explored over the past decades, the weakly interacting massive particle (WIMP) paradigm provides a particularly economical framework: a stable particle with an appropriate mass and interactions with the Standard Model (SM) can be thermally produced in the early Universe. Today DM is expected to populate the halos of galaxies, including the Milky Way, so that DM particles can pass through terrestrial detectors and potentially scatter off nuclei or electrons.

This expectation provides the basis for the direct-detection strategy and has motivated an extensive experimental
programme that targets nuclear and electronic recoils in ultra-low-background detectors on Earth. 
Experiments such as XENONnT \cite{XENON:2023cxc,XENON:2025vwd,XENON:2026gxb}, PandaX-4T \cite{PandaX-4T:2021bab,PandaX:2024qfu,PandaX:2025rrz} and LUX-ZEPLIN (LZ) \cite{LZ:2022lsv,LZ:2024zvo} have placed stringent constraints on conventional dark matter--nucleon interactions. 
Recently, LZ extended its nuclear-recoil
search window to recoil energies of approximately $270~\mathrm{keV}$,
allowing it to probe dark-matter scenarios whose recoil spectra are
substantially harder than those of conventional elastic WIMP scattering. The analysis identified an isolated nuclear-recoil-like event
at~\cite{LZHighEnergy2026}
\begin{equation}
    E^{\rmii{LZ}}_R = 248 \pm 23_{\rm stat} \pm 23_{\rm sys}~\mathrm{keV},
\end{equation}
in a region where the expected background is small. LZ reports a
background-only tension of $2.6\sigma$ after accounting for the
look-elsewhere effect, with a maximum local significance of $3.4\sigma$
among the models considered. This is certainly not a discovery of
dark matter yet, but it provides a strong motivation for studying
non-standard scenarios capable of producing nuclear recoils at such high
energies.

Among the possibilities considered in this context, inelastic dark matter
provides a particularly simple mechanism for modifying the recoil spectrum \cite{TuckerSmith:2001hy,Chang:2008gd,Schutz:2011tw}.
In an inelastic scattering process, a dark-sector state $X_1$ scatters
into a distinct state $X_2$ with a small mass splitting
$\Delta m = m_{X_2}-m_{X_1}.$
For endothermic scattering, $\Delta m>0$, part of the kinetic energy of the
incoming dark matter particle is required to produce the heavier state.
This modifies the minimum velocity required for a given nuclear recoil and
can strongly suppress low-energy scattering while enhancing the relative
importance of high-energy recoils. 

The LZ event has already motivated several recent studies of this
possibility. In particular, endothermic dark matter interpretations find
that the observed recoil energy can be naturally associated with a
dark-sector mass splitting of order a few hundred keV, with representative
solutions involving dark-matter masses at the TeV scale
\cite{Su:2026,DiMauro:2026,Freese:2026,Wang:2026ytg,Fan:2026kxx}.
Other recent analyses have compared
endothermic scattering with alternative mechanisms such as boosted dark
matter \cite{Mahapatra:2026}, while exothermic transitions between
nearly-degenerate dark states have also been considered
\cite{Baer:2026}. These studies highlight an important feature of the
high-energy event: an interpretation in terms of inelastic scattering
necessarily introduces non-trivial internal structure into the dark sector.
The dark matter particle relevant for direct detection is no longer an
\emph{isolated state}, but one member of a spectrum containing at least one
additional particle with a mass separation of order hundreds of keV.

In this work, we investigate this possibility in a leptophilic multicomponent dark-matter model. Our construction is based on a broadly studied class of simplified dark-matter models, often referred to as \(t\)-channel dark matter models~\cite{Arina:2020udz,Garny:2015wea,Arina:2025zpi,Biondini:2026dpm}, charged-mediator models~\cite{Ibarra:2015fqa} or lepton-portal dark matter~\cite{Bai:2014osa}. We use this framework to investigate the phenomenology that follows from the minimal ingredients required for dark matter to interact with SM leptons. 
The dark states are taken to be SM gauge singlets, while a vector-like fermion mediator allows a renormalisable dimension-four interaction between the dark sector and the leptons. This setup allows us to study which features arise when scalar \textit{and} vector dark states interact with right-handed Standard Model leptons through a \(t\)-channel mediator, with the aim of connecting them to the high-energy recoil event reported by LZ.

For the leptophilic dark states considered here, scattering off nuclei is induced via photon exchange, with the effective vertex generated at loop level. Models explaining the LZ event through endothermic scattering mediated by a photon have been considered previously in Refs.~\cite{Asadi:2026iot,He:2026hqz}. However, these works focus on transition dipole moments between two fermionic states and therefore lead to a different phenomenology. In our case, the leading effective operators mediating interactions between dark bosons and the photon arise at dimension six; see Refs.~\cite{Bai:2014osa,Brod:2017bsw,Kavanagh:2018xeh} for the real scalar case and Refs.~\cite{Hisano:2020qkq,Hagiwara:1986vm} for the real vector case. Interestingly, we find that the leading operators responsible for the inelastic endothermic scalar-vector transition are instead of dimension five. Moreover, the elastic interaction of the scalar state is suppressed by an additional power of the electromagnetic coupling. This hierarchy provides a natural explanation for why the inelastic transition can dominate over elastic scattering.

The same effective operator that leads to the direct detection signal also allows for the decay of the heavier dark state. Thus, only the lighter of the two is stable on cosmological scales and constitutes the dark matter today. We focus on the case where this is the scalar in the following. However, nothing prevents the opposite mass hierarchy, and investigating the vector dark matter scenario is an intriguing direction for future work.
We study the direct detection phenomenology in detail and aim to identify the parameters that lead to a recoil rate and spectrum consistent with the observation of a single high-energy recoil event by LZ. Naturally, the appeal of such a solution is significantly strengthened if the dark matter candidate can be motivated from a simple cosmological production mechanism. Therefore, we investigate the thermal history of the dark sector and determine the relic density from a freeze-out process.

The structure of this paper is as follows. In sec.~\ref{sec:model}, we briefly introduce the model before studying the set of effective operators that describe the interactions between the dark particles and the photon in sec.~\ref{sec:eff_operators}.  In sec.~\ref{sec:cosmo}, we investigate the cosmological evolution of the dark sector and determine the corresponding relic abundance. Sec.~\ref{sec:direct_det} focuses on the direct detection signal. We study the differential nuclear recoil rate for xenon-based detectors and investigate whether the parameters preferred by the LZ signal are consistent with a thermal history of the dark matter candidate. The implications of constraints from other searches are considered in sec.~\ref{sec:other_constraints}. Finally, we summarize our results in sec.~\ref{sec:conclusions}. Two appendices provide additional details for relic density computation and the velocity averages that enter into the recoil rate.
\section{Multi-component leptophilic dark matter}
\label{sec:model}
In this work, we consider a dark sector containing a real scalar $S$ and a real vector $V_\mu$, both singlets under the Standard Model (SM) gauge group. The interactions relevant for our analysis are
\begin{align}
\mathcal{L}_{\rmi{int}} \supset{}&
- \yS \, S \bar{F} P_R \ell
- \yS \, S \bar{\ell} P_L F
- \yV \, \bar{F}\gamma^\mu P_R \ell \, V_\mu
- \yV \, \bar{\ell}\gamma^\mu P_L F V_\mu \, .
\label{eq:lagrangian}
\end{align}
Here, $F$ is a vector-like charged fermion mediator, while $\ell$ denotes a right-handed SM charged lepton. Gauge invariance then fixes the hypercharge of the mediator to $Y_F=Y_\ell=-1$. The dark sector particles are taken to be odd under a discrete $\mathcal{Z}_2$ symmetry, while all SM fields are even. This symmetry stabilizes the lighter of the two dark bosons.
We take $\yS$ and $\yV$ to be  real.

The interactions in Eq.~\eqref{eq:lagrangian} belong to the well-studied class of $t$-channel lepton-portal models, in which a dark state couples to SM leptons through a charged mediator. The latter can also be referred to as a vector-like lepton. Such models have been considered separately for scalar, fermion and vector dark matter, with the corresponding phenomenology studied under the assumption that only one of these states constitutes the dark matter \cite{Garny:2015wea,Arina:2020udz,Yu:2014pra}. Here we combine the two bosonic possibilities within a single dark sector. More generally, our construction provides a simple framework in which bosonic dark states of different spin, namely a scalar and a vector, can couple to right-handed SM leptons through the same vector-like charged fermionic mediator. The usual scalar- and vector-dark matter scenarios are recovered as limiting cases in which one of the two dark states is absent or phenomenologically irrelevant. This more general setup allows both states to be populated during thermal freeze-out and to participate in the coupled cosmological evolution of the dark sector. Most importantly for this work, they allow for an inelastic transition when scattering off nuclei. 

Regarding the flavour structure of the interactions in Eq.~\eqref{eq:lagrangian}, we consider two possibilities. In the first, the dark bosons $S$ and $V_\mu$ couple to the same SM lepton flavour, with only one flavour being coupled at a time. This choice is commonly adopted in phenomenological studies of simplified dark-matter models \cite{Garny:2015wea,Bollig:2021psb,Arina:2020udz,Biondini:2019int,Arina:2025zpi}. In the second, the dark bosons couple to all three lepton flavours through diagonal Yukawa couplings of equal strength, with three identical copies of the fermionic mediator, $F^a$, where $a=e,\mu,\tau$; see, e.g., \cite{Kopp:2014tsa,Biondini:2026dpm}. More general constructions with non-trivial flavour structures, including off-diagonal couplings between different lepton flavours, have also been considered; such couplings can be subject to stringent lepton-flavour-violation (LFV) constraints; see, e.g. \cite{Kopp:2014tsa,Arina:2025zpi,Belfatto:2025ids,Biondini:2026ryb}.

We assume that the scalar is the lighter state,
\begin{equation}
m_V > m_S,
\qquad
\Delta m_{\scriptscriptstyle SV} \equiv m_V-m_S >0.
\label{eq:mass_splitting}
\end{equation}
The mass splitting of interest is parametrically small and is motivated by the high-energy nuclear-recoil event reported by the LZ experiment \cite{LZ:2026axp}. We therefore consider
\begin{equation}
\Delta m_{\scriptscriptstyle SV} \sim {\cal O}(10^2)~{\rm keV},
\label{eq:delta_m}
\end{equation}
while the overall dark-sector mass is taken to lie from a few hundred GeV to the TeV scale, as suggested by the thermal relic-density requirement with $\yS,\yV$ of $\mathcal{O}(1)$. The heavier vector state is unstable and eventually decays into the lighter scalar state. Nevertheless, it will be populated in the early Universe and participate in the thermal freeze-out of the dark sector.  For later use, we also define the positive mass splittings between the fermionic mediator and the dark bosons as
\begin{align}
    \DmFS \equiv m_F -m_S \, , \quad  \DmFV \equiv m_F -m_V \, ,
\end{align}
and the relative mass splittings as $\delta m_{ij} \equiv \Delta m_{ij}/m_j$, which are dimensionless quantities.

We assume that possible additional portals to the SM are sufficiently suppressed that they do not affect the phenomenology considered here. In particular, the scalar can interact with the Higgs through the usual Higgs-portal operator, $-\lambda_{\scriptscriptstyle SH} S^2 H^\dagger H$, and a similar operator can be envisaged for the dark vector, $-\lambda_{\scriptscriptstyle VH} V^\mu V_\mu H^\dagger H$, whereas a vector portal through kinetic mixing is instead forbidden by the $\mathcal{Z}_2$ symmetry of the dark sector.
We take these interactions to be subdominant with respect to the lepton-portal couplings $g$ and $y$, allowing us to isolate the phenomenology associated with the latter. This assumption is commonly adopted in simplified $t$-channel dark-matter models~\cite{Arina:2020udz,Garny:2015wea,Arina:2025zpi}.

\section{Effective dark-sector SM operators}
\label{sec:eff_operators}
For scalar dark matter with a lepton portal, the coupling to nuclei is absent at tree level and direct detection is instead mediated by loop-induced interactions with photons. In particular, the leading contributions can be described in terms of dimension-six Rayleigh operators involving two electromagnetic field strengths \cite{Bai:2014osa,Brod:2017bsw,Kavanagh:2018xeh}. Analogous fermion-portal constructions have also been studied for vector dark matter, including scenarios in which a vector dark state interacts with SM fermions through a heavy fermionic mediator~\cite{Yu:2014pra}. Similar to the scalar dark matter case, the leading interaction between the dark vector and the photon is also found to be of dimension-six. Interestingly, the leading contribution to the inelastic scattering already arises at dimension five. Thus, one expects a significant enhancement of the inelastic scattering rate compare with the elastic one.

The same effective interactions between the dark sector and photons have an impact on the thermal history of the model. Since our working assumption is $m_S<m_V$,  the lighter scalar constitutes the stable dark matter state at late times, while the heavier vector participates in freeze-out and can undergo a transition into the scalar state through its effective electromagnetic interactions. 
The effective dark-sector SM operators are derived below and provide the basis for the direct-detection analysis in section~\ref{sec:direct_det}, as well as govern the lifetime of the heavier vector state and, therefore, enter the cosmological analysis of section~\ref{sec:cosmo}.

The leading interactions between the SM-singlet states $S$ and $V_\mu$ and the SM can be described by dimension-five effective operators. Restricting to operators involving photons and leptons a convenient basis at dimension five can be written as
\begin{align}
\mathcal{L}_{\rmi{eff}}^{(5)} \supset{}&
\frac{a_\gamma}{m_S} S V_{\mu\nu} F^{\mu\nu}
+\frac{b_\gamma}{m_S} S V_\mu \partial_\nu F^{\mu\nu}
+\frac{c_{\ell}}{m_S}
S V_\mu \, \bar{\ell}\gamma^\mu\ell 
+\frac{c_{\scriptscriptstyle L}}{m_S}
S V_\mu \, \bar{L}\gamma^\mu L  \, ,
\label{eq:dim5_operators}
\end{align}
where $F^{\mu\nu}$ denotes the electromagnetic field-strength tensor, $\ell$ and $L$ denote the right- and left-handed charged-lepton fields, respectively, and flavour indices are suppressed.
While CP-odd operators also appear at dimension five, they are not generated by the UV-completion considered here and have been omitted them. In addition, we have dropped analogous operators involving the other SM gauge bosons. Their contributions are subdominant for the nuclear-recoil phenomenology considered here. In particular, the $Z$-mediated contribution is suppressed by the heavy-boson propagator at the momentum transfers relevant for direct detection.

\begin{figure}
    \centering
    \includegraphics[width=0.95\linewidth]{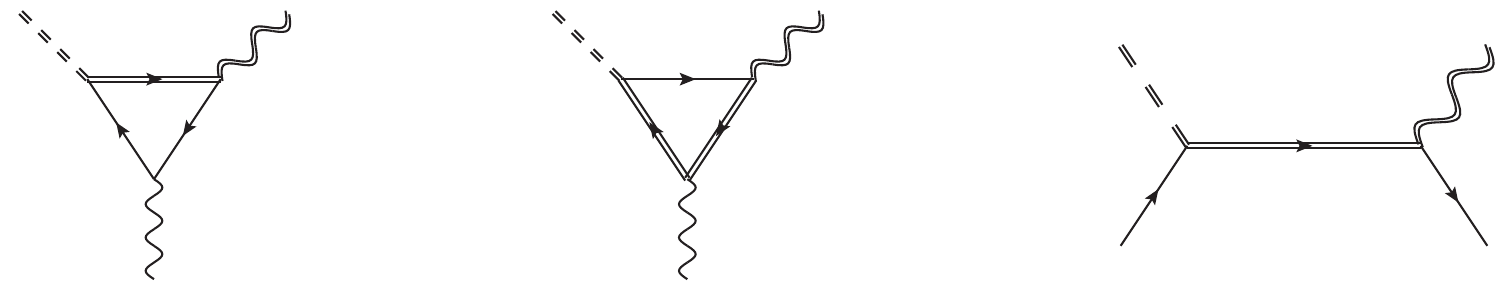}
    \caption{Exemplary one-loop and tree-level diagrams contributing to the matching of the dimension-five operators in the effective Lagrangian of eq.~\eqref{eq:dim5_operators}. SM right-handed leptons are represented by solid lines, while photons are represented by wavy lines. Dark-sector vector, scalar and fermion fields are instead represented by wavy, dashed and solid double lines, respectively.}
    \label{fig:matching_diagrams}
\end{figure}
The matching coefficients of the effective operators involving the photon arise at one-loop level, since the dark states $S$ and $V_\mu$ are neutral under electromagnetism. Representative one-loop diagrams contributing to the matching are shown in Fig.~\ref{fig:matching_diagrams}. We perform the matching through an automated computational pipeline based on
\texttt{FeynRules}~\cite{Alloul:2013bka},
\texttt{FeynArts}~\cite{Hahn:2000kx},
\texttt{FeynCalc}~\cite{Shtabovenko:2020gxv},
and \texttt{Package-X}~\cite{Patel:2015tea}. Moreover, we have independently checked the expressions for the loop amplitudes. 

Let us start from the one-loop induced effective interactions among the photon and the dark scalar and dark vector. The extraction of the matching coefficients depends on the lepton flavour. As discussed in refs.~\cite{Kopp:2014tsa,Biondini:2026dpm}, the distinction originates from the relative size of the typical momentum transfer and the mass of the SM lepton running in the loop. For non-relativistic nuclear recoils the momentum transfer is related to the recoil energy through \(|\bm{q}|^2=2m_{N} E_R\). 
Taking $E_R \sim 250 $ keV, as observed by LZ, one finds \(\sqrt{-q^2} \simeq |\bm{q}|\simeq250\) MeV 
for the typical momentum transfers. Thus we take the following hierarchy of scales to hold for the tau lepton
\begin{equation}
    q^2 \ll m_{\tau}^2 \ll m_S^2 \simeq m_V^2  < m_F^2 \, .
    \label{eq:hierachy_heavy_leptons}
\end{equation}
The small-momentum-transfer limit can therefore be taken safely when expanding the amplitude, allowing us to neglect terms suppressed by powers of $q^2/m_{\tau}^2$, which would then correspond to higher-dimensional effective operators.
The result for the matching coefficients then read
\begin{align}
a_\gamma^{(\tau)}
={}&
\frac{e \yS \yV}
{16 \pi^2 \rho^{\frac{5}{2}}}
\Bigg\{
(1+\sqrt{r})\rho
\left(-2+4\sqrt{r}-2r+\rho\right)
\nonumber\\
&
+(-1+\sqrt{r})
\left[1+r^2-\rho-\sqrt{r}\rho-r(2+\rho)\right]
\log r
\nonumber\\
&
-\frac{2(1+\sqrt{r})}{\sqrt{\lambda(1,r,\rho)}}
\Big[
-2r^{5/2}+r^3+(1-\rho)^2
-r^2(1+2\rho) +r^{3/2}(4+3\rho)
\nonumber\\
&
-\sqrt{r}(2-3\rho+\rho^2)
+r(-1-2\rho+\rho^2)
\Big]
\log\left(
\frac{1+r-\rho+\sqrt{\lambda(1,r,\rho)}}
{2\sqrt{r}}
\right)
\Bigg\} \, ,
\end{align}
and 
\begin{align}
b_\gamma^{(\tau)}
={}&
\frac{e \yS \yV}
{24 \pi^2 \rho^{\frac{5}{2}}}
\Bigg\{
4(-1+\sqrt{r})^2(1+\sqrt{r})\rho
\nonumber\\
&
- (1-\sqrt{r})
\left[-2-2r^2+\rho+\rho^2+r(4+\rho)\right]
\log r
\nonumber\\
&
+\frac{2(1+\sqrt{r})}{\sqrt{\lambda(1,r,\rho)}}
\Big[
2-4r^{5/2}+2r^3-3\rho-\rho^3 -2r(1+3\rho)
\nonumber\\
&
-r^2(2+3\rho)
+\sqrt{r}(-4+6\rho)
+r^{3/2}(8+6\rho)
\Big]
\log\left(
\frac{1+r-\rho+\sqrt{\lambda(1,r,\rho)}}
{2\sqrt{r}}
\right)
\Bigg\}.
\end{align}
where the dimensionless parameters that control the size of the loop function are expressed in terms of the mass ratios
\begin{align}
 \rho=\frac{m_S^2}{m_F^2} \, , \quad    
r=\frac{m_\ell^2}{m_F^2} \, .
\label{ratio_mass_scales}
\end{align}
The mass splitting between the dark boson, $\Delta m_{\scriptscriptstyle VS}$ is much smaller than $q^2$ and it is set to zero in the matching procedure. Accordingly there is no corresponding mass ratio for $m_V^2/m_F^2$ appearing in the expressions for $a^{(\tau)}_\gamma$ and $b^{(\tau)}_\gamma$.

Instructive limits for the matching coefficients can be exacted at leading order for $\rho<1$ and $r \ll 1$. The latter condition is always well satisfied in the parameter range of interest in our work, with $m_S$ being larger than $\mathcal{O}(10^2)$ GeV. The leading contributions are
\begin{align}
    a^{(\tau)}_\gamma \simeq  - e \yS \yV \frac{\sqrt{\rho}}{48 \pi^2} \, , \quad  b^{(\tau)}_\gamma \simeq e \yS \yV \frac{\sqrt{\rho} \left( 5 + 6 \ln r\right)}{72 \pi^2} \, .
    \label{eq_a_b_heavy_limit}
\end{align}
As can be seen, $b^{(\tau)}_\gamma$ is logarithmically enhanced but otherwise of similar order as $a^{(\tau)}_\gamma$.

The matching procedure is slightly different for the muon and  electron. In this case the arrangements of the scales is reasonably represented by
\begin{align}
         m_{e}^2, m_{\mu}^2  \ll q^2 \ll m_S^2 \simeq m_V^2  < m_F^2 \, .
    \label{eq:hierachy_light_leptons}
\end{align}
and, therefore, the lepton masses are set to zero in the expansion of the one-loop amplitude in the matching. The infrared sensitivity, that appears only for the coefficient $b_\gamma$ is now encoded in the ration 
\begin{equation}
    \xi \equiv - \frac{q^2}{m_F^2} \, , 
\end{equation}
and the coefficients read 
\begin{align}
    & a^{(e,\mu)}_\gamma = \frac{e \yS \yV}{16 \pi^2 \rho^{\frac{5}{2}}} \left[ \rho^2 - 2\rho - 2(1-\rho) \ln (1-\rho) \right] \, ,
    \\
    & b^{(e,\mu)}_\gamma = \frac{e \yS \yV}{36\pi^2\rho^{5/2}(1-\rho)}
\Bigg[
6\rho - 6\rho^2 - 5\rho^3
+ (6 - 9\rho - 3\rho^3)\ln(1-\rho)
+ 3\rho^3\ln\xi
\Bigg].
\end{align}
The corresponding limits of the light-lepton operator coefficient $b_{\gamma}^{(e,\mu)}$ for $\rho <1$ and $\xi \simeq 2 E_R m_A /m_F^2 \ll 1$ is 
\begin{equation}
     b^{(e,\mu)}_\gamma  \simeq  e \yS \yV \frac{\sqrt{\rho}(-5 + 6 \ln \xi)}{72 \pi^2} \, ,
\end{equation}
whereas one finds the same limit as in eq.~\eqref{eq_a_b_heavy_limit} for $a_{\gamma}^{(e,\mu)}$. Hence, the $a_\gamma$ coefficient is practically the same for all flavours for the range of splitting between the dark scalar and the fermionic mediator considered in our work.

In contrast to the photon operators, the operators involving SM leptons are generated already at tree level through the exchange of the heavy fermion mediator, as illustrated by the rightmost diagram in Fig.~\ref{fig:matching_diagrams}. For the flavour structure assumed in Eq.~\eqref{eq:lagrangian}, the matching coefficients are
\begin{align}
c_{\scriptscriptstyle L} &= 0,
\qquad
c_{\ell}
= \frac{2 \yS\yV m_S}{m_F}\,.
\label{eq:matching_lepton}
\end{align}
The vanishing of $c_{\scriptscriptstyle L}$ follows directly from the purely right-handed lepton coupling in the underlying Lagrangian. 
If the dark sector is coupled to electrons, these lepton-portal interactions can give rise to an electron-recoil signal. 

We note in passing that the operators corresponding to the matching coefficients $c_\ell$ and $b_\gamma$ in the effective Lagrangian \eqref{eq:dim5_operators} mix at one loop. Since $c_\ell$ is generated at tree level, this mixing can potentially give rise to sizeable effects. By contrast, at one loop and with a single insertion of the operators in the effective Lagrangian, $a_\gamma$ does not undergo any mixing. We refrain from performing a one-loop analysis and running of $b_\gamma$ given that the contribution of this operator to the nuclear scattering cross section will turn out to be negligible; see sec.~\ref{sec:direct_det}.

\section{Cosmology of the dark sector}
\label{sec:cosmo}

In this section, we investigate the cosmology of the dark sector. Although our primary goal is to assess the possible interpretation of the high-energy recoil event reported by the LZ experiment, it is important to establish the corresponding thermal history and verify that the model can reproduce the observed dark matter abundance. Therefore, we aim to identify benchmark values of the masses and couplings that are compatible both with the LZ event and with the observed dark matter abundance $\Omega_{\rm DM} h^2 = 0.1200 \pm 0.0012$ measured by Planck \cite{Planck:2018vyg}. In addition, we investigate the lifetime of the heavier vector state to assess whether its decay can have consequences for cosmological observables.

\subsection{Dark matter freeze-out}
\label{sec:freeze_out} 
At the characteristic timescales associated with thermal freeze-out for dark-sector masses of $\mathcal{O}(1~{\rm TeV})$, the heavier vector state is effectively stable. Thus  the vector participates in the processes maintaining chemical equilibrium between the dark sector and the SM thermal bath at $T \gg m_S$, and which is then progressively lost for $M_S/T \simeq 25$.\footnote{A simple estimate of the relevant timescale can be obtained by comparing the interaction rate with the Hubble expansion rate around chemical freeze-out. Chemical equilibrium is maintained when $\Gamma_{\rm int}\gtrsim H$, where, schematically, $\Gamma_{\rm int}\sim n_S^{\rm eq}\langle\sigma v_{\rm rel}\rangle$. During radiation domination, the Hubble rate is $H=\sqrt{4\pi^3g_{\rm eff}/45}\,T^2/m_{\rm Pl}$. Taking $T\simeq m_S/25\simeq40\,\mathrm{GeV}$ for $m_S\simeq1\,\mathrm{TeV}$ and $g_{\rm eff}\simeq100$, we obtain a Hubble timescale $t_{\rmii{fo}}\equiv H^{-1}\simeq3\times10^{-10}\,\mathrm{s}$. The vector decay can therefore be neglected during freeze-out provided that $\Gamma_V\ll H(T_{\rmii{fo}})$, or equivalently $\tau_V\gg t_{\rmii{fo}}\simeq3\times10^{-10}\,\mathrm{s}$.} In addition, the vector-like fermion mediator could also contribute to the thermal dynamics (see e.g.~refs.\cite{Biondini:2026dpm,Biondini:2019int} for lepto-philic and quark-philic scalar dark matter). A general treatment therefore involves three coannihilating species, $S$, $V$, and $F$. The relevance of states heavier than the stable dark matter particle is controlled by their mass splittings relative to $m_S$, $\dmFS$  and $\dmVS$, through the corresponding Boltzmann suppression of their equilibrium abundances~\cite{Griest:1990kh,Edsjo:1997bg}.

In the present scenario, the scalar and vector states are highly degenerate, with a mass splitting of order $\Delta m_{\scriptscriptstyle SV}\sim\mathcal{O}(10^2)~{\rm keV}$ whereas $m_S \sim \mathcal{O}(100)$ GeV, as motivated by the direct-detection phenomenology, cf.~eq.~\eqref{eq:delta_m}. We therefore always need to retain both bosonic states in the freeze-out calculation. On the other hand, the relevance of the vector-like fermion depends on the relative splitting with the bosonic states.  For mass splittings satisfying
\begin{equation}
\frac{\DmFS}{m_S}
\simeq
\frac{\DmFV}{m_V}
\gtrsim 0.3,
\label{eq:splitting_F_cond}
\end{equation}
the equilibrium abundance of the mediator is already sufficiently Boltzmann suppressed at freeze-out that its contribution to the effective annihilation rate is negligible \cite{Griest:1990kh,Baker:2015qna}. We can therefore describe the relevant freeze-out dynamics in terms of the two bosonic states, $S$ and $V$. Whenever the splitting with the vector-like lepton is smaller, we lift this simplifying assumption and include the corresponding coannihilation processes with the fermionic mediator. In this case, the three species contribute to the determination of the DM energy density.

In the following we describe the scenario with $S$ and $V$ driving the DM relic abundance, whereas coannihilations with the fermion are relegated to Appx.~\ref{sec:App_B}. 
When the condition in eq.~\eqref{eq:splitting_F_cond} holds, the effective annihilation rate consequently receives contributions from the scalar and vector annihilation processes,
$S S \rightarrow \ell\bar{\ell}$, $V V \rightarrow \ell\bar{\ell}$,
as well as from the coannihilation process
$S V \rightarrow \ell\bar{\ell}$.
As long as the conversion processes between the two dark states are sufficiently rapid to maintain chemical equilibrium, the evolution of the total dark-sector number density can be described by an effective Boltzmann equation~\cite{Griest:1990kh,Edsjo:1997bg},
\begin{equation}
\frac{d n}{dt} + 3H n
=
-\langle \sigma_{\rm eff} v_{\rm rel} \rangle
\left(n^2-n_{\rm eq}^2\right),
\label{BE_start}
\end{equation}
where
$ n=n_{\scriptscriptstyle S}+n_{\scriptscriptstyle V}$
denotes the total number density of the two bosonic dark-sector states, and $H$ is the Hubble rate. 
The effective annihilation cross section is given by
\begin{equation}
\sigma_{\rm eff} v_{\rm rel}
=\frac{1}{\left(\sum_k n_k^{\rm eq}\right)^2}
\sum_{i,j}
n_i^{\rm eq}
n_j^{\rm eq}
\sigma_{ij}v_{\rm rel},
\label{eq:sigma_eff_general}
\end{equation}
where the indices $i,j$ run over the dark-sector species retained in the freeze-out calculation, here $i,j=S,V$. The equilibrium densities encode the Boltzmann suppression of the heavier state and therefore determine the relative weight of the different annihilation and coannihilation channels. 
Due to the vector-scalar splittings relevant to our work, $\mathcal{O}(100)$ keV and dark matter mass $m_S$ larger than $\mathcal{O}(100)$ GeV, the effective cross section can be greatly simplified as the exponential factors are practically unity during freeze-out 
\begin{align}
      \sigma_{\textrm{eff}} \vrel &= \frac{ \sigma_{\scriptscriptstyle S S} \vrel + 2 \frac{g_{\scriptscriptstyle V}}{g_{\scriptscriptstyle S}}  \sigma_{\scriptscriptstyle S V} \vrel   +  \frac{g_V^2}{g_S^2}  \sigma_{\scriptscriptstyle VV} \vrel }{\left( 1+ \frac{ g_{\scriptscriptstyle V}}{g_{\scriptscriptstyle S}}  \right)^2} \, ,
\label{eff_Xsection_coann_real_S_v2}
\end{align}
where the internal particle degrees of freedom are $g_S=1$ and $g_V=3$, which account for the massive vector polarization.

The cross sections for the pure-scalar and pure-vector cases are already known in the literature, see e.g.~\cite{Bai:2014osa,Yu:2014pra}, whereas we compute here the cross section for the \(SV \to \ell \bar{\ell}\) process. At leading order in the velocity expansion, they read
\begin{align}
    \sigma_{\scriptscriptstyle SS} \vrel & = \frac{\yS^4 m_S^6}{60\pi\,(m_S^2+m_F^2)^4}\vrel^4\, ,
    \label{SS_ll}
    \\
     \sigma_{\scriptscriptstyle SV} \vrel &= \dfrac{\yS^2 \yV^2 m_F^2   (m_S +m_V)^2}{12 \pi\, m_S m_V\, (m_F^2 +m_S m_V)^2 } \simeq  \dfrac{\yS^2 \yV^2 m_F^2   }{3 \pi\,  (m_F^2 +m_S^2)^2 } \, ,
     \label{SV_ll}
    \\
    \sigma_{\scriptscriptstyle VV} \vrel  &= \frac{2m_V^2 \yV^4}{9 \pi \,(m_V^2+m_F^2)^2} \simeq \frac{2m_S^2 \yV^4}{9 \pi\, (m_F^2+m_S^2)^2}\, .
      \label{VV_ll}
\end{align}
The cross sections for \(SV \to \ell \bar{\ell}\) and \(VV \to \ell \bar{\ell}\) can be well approximated by the expressions in the second equalities of eqs.~\eqref{SV_ll} and \eqref{VV_ll}, respectively, by neglecting corrections of order \(\Delta m_{\scriptscriptstyle SV}/m_S \ll 1\). The dependence of the \(SV\) annihilation cross section on the mediator mass differs from that of the \(SS\) and \(VV\) channels. In this case, the chiral projectors select the fermion-mass term in the \(t\)- and u-channel diagrams, thereby modifying the power of \(m_F\) in the cross section. This behaviour is also connected to the dimension-five operator involving the dark bosonic fields and the right-handed leptons, which is generated upon integrating out the mediator (see sec.~\ref{sec:eff_operators}).

The effective cross section can then be recast into a convenient form, which allows us to identify from the outset the dominant contributions. Using the $\rho$ as defined in eq.~\eqref{ratio_mass_scales},
we can write the thermally averaged effective cross section in eq.~\eqref{eff_Xsection_coann_real_S_v2} as
\begin{align}
    \langle \sigma_{\rmi{eff}} \vrel \rangle = \frac{\yS^2 \yV^2}{\pi m_F^2}  \left[ \frac{\yS^2}{\yV^2}\frac{\rho^3}{(1+\rho)^4 z^2} + \frac{g_{\scriptscriptstyle V}}{g_{\scriptscriptstyle S}}\frac{2}{3(1+\rho)^2} + \frac{\yV^2}{\yS^2} \frac{g_{\scriptscriptstyle V}^2}{g_{\scriptscriptstyle S}^2}\frac{2 \rho}{9(1+\rho)^2}\right] \frac{1}{\left( 1+ \frac{ g_{\scriptscriptstyle V}}{g_{\scriptscriptstyle S}}  \right)^2} \, ,
    \label{eff_cross_rho}
\end{align}
where $z = m_S/T$ is the usual dimensionless variable used to track the dark matter abundance in the Boltzmann equation. 

An inspection of eq.~\eqref{eff_cross_rho} suggests that, for equal couplings, $\yS = \yV$, the relic density is mainly determined by the coannihilation processes $SV \to \ell \bar{\ell}$ and $VV \to \ell \bar{\ell}$, since scalar-pair annihilation is $d$-wave suppressed (with $z \simeq 25$ at freeze-out). Moreover, in the heavy-mediator limit, $\rho \ll 1$, the annihilation process $SV \to \ell \bar{\ell}$ provides the dominant contribution. In addition to the equal-coupling scenario, we consider two further possibilities for the relative size of the couplings, in which either $\yS$ or $\yV$ is fixed to unity, while the other coupling is allowed to vary in the range $[0.1,\sqrt{4\pi}]$.
\begin{figure}
    \centering
    \includegraphics[width=0.48\linewidth]{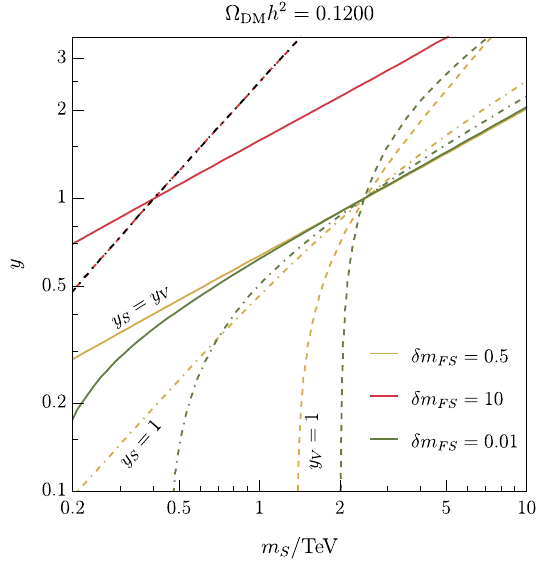}
    \includegraphics[width=0.48\linewidth]{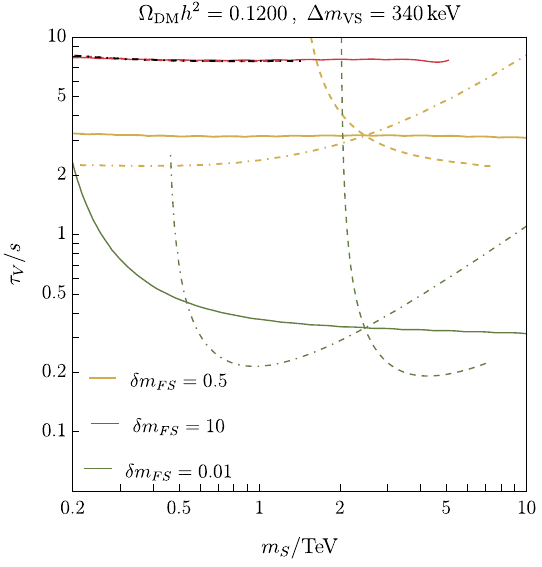}
    \caption{(Left) Contours corresponding to the observed relic density in the $(m_S,y)$ plane. Here, $y$ generically denotes either $\yS$ or $\yV$, depending on the coupling-assignment scenario. The solid, dashed and dot-dashed lines correspond, respectively, to $\yS=\yV$, $\yV=1$ (with $\yS$ varied) and $\yS=1$ (with $\yV$ varied). The red, ochre and olive lines correspond to $\delta m_{\scriptscriptstyle FS}=10$, $0.5$ and $0.01$, respectively. For $\delta m_{\scriptscriptstyle FS}=10$, some of the lines are degenerate; we use black for these to improve readability. (Right) Dark-vector lifetime as a function of the dark matter mass $m_S$ for the same three coupling assignments and values of $\delta m_{\scriptscriptstyle FS}$ as in the left panel. The scalar-vector mass splitting is fixed to $\DmVS=340$ keV.
    }
    \label{fig:relic_density}
\end{figure}

In figure~\ref{fig:relic_density}, we explore the three coupling assignments for three benchmark values of the dark matter--mediator relative mass splitting, $\delta m_{\scriptscriptstyle FS} = \DmFS/m_S = 0.01$ (olive), 0.5 (ochre), and 10 (red). For each choice, the solid, dashed and dot-dashed black lines show, respectively, the contours corresponding to the observed relic density in the $(m_S,y)$ plane for $\yS = \yV$, $\yV=1$ (with $\yS$ varied) and $\yS=1$ (with $\yV$ varied).
By construction all lines for a given mass hierarchy meet at $y=1$, which happens at $m_S \simeq 400$ GeV for $\delta m_{\scriptscriptstyle FS}=10$ and $m_S \simeq 2000$ GeV for $\delta m_{\scriptscriptstyle FS}=0.5$ and $\delta m_{\scriptscriptstyle FS}=0.01$. Qualitatively, the further behaviour of these curves can be understood as follows. Let us first consider the heavy-mediator limit, in which the annihilation cross section is dominated by the mixed-channel $SV$ due to the lower $1/m_F$ suppression. Thus one finds $\langle \sigma_{\rmi{eff}} \vrel \rangle \propto \yS^2 \yV^2/m_F^2$ and the two scenarios with one coupling fixed to unity and the other varied become degenerate, as reflected by the overlapping dashed and dot-dashed curves. In contrast, when the two couplings are equal, the effective cross section scales as $y^4$, rather than as $y^2$ when one coupling is fixed to unity. Consequently, larger values of the common Yukawa coupling are required to obtain the same annihilation rate until the coupling reaches unity and vice versa at larger couplings. 
For a more compressed dark matter--mediator mass spectrum some features remain while others change. As before, all lines intersect at $y=1$ and the fixed coupling scenarios require smaller (larger) couplings for masses below (above) the intersection. However, the degeneracy between the two scenarios with one Yukawa coupling fixed to unity is lifted as the other annihilation channels start to contribute significantly. 
In particular, the $d$-wave suppression requires a larger value of $\yV$ to reproduce the observed relic density, whereas fixing $\yV=1$ leads to a significant minimal annihilation cross section from $VV$ alone that exceeds the one required by thermal freeze-out for $m_S\lesssim 1.5$ TeV for $\delta m_{\scriptscriptstyle FS}=0.5$. This effect gets exacerbated for $\delta m_{\scriptscriptstyle FS}=0.01$ where coannihilation processes involving the fermion add to the effective cross section. This pushes the minimal mass to about $2$ TeV for the $y_V=1$ case.

\subsection{Lifetime of the dark vector}
\label{sec:vector_decay}

 The dimension-five operators outlined in eq.~\eqref{eq:dim5_operators}  contribute to the transition between the dark vector and lighter states, the dark scalar and SM particles. For the small mass splittings considered here, the vector can only decay through the two-body process
\begin{equation}
V \rightarrow S + \gamma ,
\end{equation}
with the photon produced on shell.  The operator
$S V_\mu \partial_\nu F^{\mu\nu}$ does not contribute to the decay amplitude, since
$\partial_\nu F^{\mu\nu}=0$ for an on-shell photon. The decay is controlled by the same dimension-five operator, $S V_{\mu \nu} F^{\mu \nu}$, that induces the dominant contribution to the nuclear recoil, see sec.~\ref{sec:diff_nuclear_recoil}.

The corresponding decay width is found to be
\begin{align}
\Gamma_V
&=
\frac{a_\gamma^2}{24\pi}
\frac{(m_V+m_S)^3(m_V-m_S)^3}
{m_S^2 m_V^3} \simeq
\frac{a_\gamma^2}{3\pi}
\frac{\DmVS^3}{m_S^2}
\, ,
\label{eq:vector_decay_width_approx}
\end{align}
where we used
$m_V+m_S\simeq 2m_S$ and neglected very small $\DmVS/m_S$ corrections. The corresponding lifetime, $\tau_V = \Gamma_V^{-1}$ sets the characteristic timescale of the vector decay. 
At variance with the relic density the vector lifetime is highly sensitive to the mass splitting $\DmVS$. 

A convenient approximation for the vector lifetime can be obtained by approximating $\alpha_\gamma$ for $\dmFS \gtrsim 2$ (or equivalently $\rho \lesssim 0.1$) by its leading contribution in eq.~\eqref{eq_a_b_heavy_limit}. This approximation gives an error of 5\% or less, which decreases for increasing scalar-mediator mass splittings. In this case, the lifetime reads
\begin{align}
    \tau_V \simeq \frac{1728 \, \pi^4}{\alpha_{\rmii{QED}} \yS^2 \yV^2} \frac{m_S^2}{\rho \, \DmVS^3} \, .
\end{align}
Using the values preferred by a thermal relic, we find typical lifetimes of the vector state of $\mathcal{O}(1)\,\mathrm{s}$; see the right panel of Fig.~\ref{fig:relic_density} for an illustration. This corresponds to a very long time separation on the timescales relevant for the LZ detector, and we therefore expect the vector to leave the detector before decaying. Even if a lingering vector were to decay inside the detector, the decay would not be correlated with the original recoil event by the collaboration because of the large time separation.

One might wonder whether the comparatively long lifetime of the vector could have cosmological implications. In particular, Big Bang Nucleosynthesis (BBN) comes to mind, given that the lifetime is comparable to the age of the Universe when this process is active. However, simple estimates demonstrate that these decays are harmless from a cosmological point of view. First, the energy of the emitted photon is set by the mass splitting, which is a few hundred keV in the part of parameter space of interest here (see sec.~\ref{sec:direct_det}). This is substantially below the deuterium binding energy of  2.2 MeV. Since this is the lowest binding energy among the isotopes relevant for BBN, the photons produced in the decays are well below the thresholds for nuclear photodissociation and therefore cannot directly affect the BBN reaction network.

In addition, BBN could in principle be affected by entropy production or changes in the Hubble rate. However, these effects are exceedingly small and do not lead to consequences that are anywhere near detectable. This can be seen by considering the ratio \(n_{\rm DM}/n_\gamma\), which is conserved after electron-positron annihilation. For a DM candidate with the abundance measured by Planck, \(n_{\rm DM}/n_\gamma \sim 3 \times 10^{-9} (\text{GeV}/m_S)\) in the late Universe. Thus, the decays add only about one photon per \(10^{11}\) photons in the bath for dark matter masses of \(\mathcal{O}(10^3)\) GeV. At the relevant times, i.e.~a few seconds after the Big Bang, the plasma temperature is in the range of several hundred keV. The released photons are therefore not exceptionally energetic compared to the thermal plasma, and the resulting entropy release and changes in the composition of the energy density are completely negligible.

The estimate outlined above is actually a conservative upper bound since the relative abundance of the two dark matter components is set by the freeze-out of their interconversion processes. If the vector fraction is further reduced by the time of its decay, the estimate above, which assumes the full dark matter abundance is available for decay, is accordingly reduced. 

\section{Inelastic nuclear scattering}
\label{sec:direct_det}

In this section, we carry out the central analysis of this work. We first discuss the effective operators that describe the interactions between the scalar and vector dark states and the SM, focusing in particular on their couplings to photons and leptons. The photon interactions are especially relevant for direct detection, as they provide the leading contribution to scattering off the nuclear constituents and are therefore the main focus of our analysis. We then use the dimension-five operators and their matching coefficients to compute the differential nuclear-recoil rate  and assess the number of events expected in a LZ-like detector.

\subsection{Differential recoil rate}
\label{sec:diff_nuclear_recoil}

In this section, we collect the expressions for the differential nuclear recoil cross section as a function of the recoil energy for the inelastic process
\begin{align}
S(p)+N(k)\to V(p')+N(k') \, .
\end{align}
The frame independent differential cross section for a $2\to2$ scattering process between the dark bosons and a target nucleon is given by
\begin{align}
    \frac{d \sigma_{\scriptscriptstyle SN\to VN}}{dt}=\frac{|\mathcal{M}(SN \rightarrow VN)|^2}{16 \pi \lambda(s,m_S^2,m_N^2)}
\end{align}
where $\lambda$ is the K\"all\'en function and $m_N$ the nucleon mass while $s$ and $t$ denote the usual Mandelstam variables.
Taking the non-relativistic limit in the laboratory frame leads to
\begin{align}
\frac{d\sigma_{\scriptscriptstyle SN\to VN}}{dE_R}
&=
\frac{|\mathcal{M}(SN\to VN)|^2}
{32\pi \, m_S^2 \, m_N \, v^2} \, ,
\label{inelastic_DM_N}
\end{align}
where $v\equiv|\bm{v}_{\scriptscriptstyle S}|$ and we used $t=-|\bm{q}|^2\approx -2m_N E_R$. 
The squared matrix element is evaluated at the scattering angle fixed by the inelastic kinematic constraint,
\begin{align}
\cos\theta
&=
\frac{|\bm{q}|}{2\mu_{SN}v}
+
\frac{\Delta m_{SV}}{|\bm{q}|v} \, ,
\label{kin_mat_element}
\end{align}
where $\mu_{SN}$ is the dark matter--nucleon reduced mass. 
The kinematic constraint $|\cos\theta|\leq1$ can equivalently be expressed in terms of the minimum incoming velocity required to produce a given recoil energy,
\begin{align}
v_{\min}(E_R)
&=
\frac{1}{\sqrt{2m_NE_R}}
\left(
\frac{m_NE_R}{\mu_{SN}}
+\DmVS
\right) \, \
\label{v_min_eq}
\end{align}
which recovers the elastic limit for $\DmVS \to0$.

For scattering on a nucleus, the same expressions apply upon replacing the nucleon mass and reduced mass with their nuclear counterparts, $m_N\to m_A$ and $\mu_{SN}\to\mu_{SA}$. The matrix element must also be corrected to account for the nuclear charge distribution through a nuclear form factor $F$ which accounts for the  coupling of the photon to the nuclear charge distribution.
The squared matrix element receives contributions from both operators in eq.~\eqref{eq:dim5_operators}, as well as from their interference. Retaining the leading terms in each contribution, we obtain
\begin{align}
   | \mathcal{M}_{\scriptscriptstyle S A \to V A}|^2 &= 4 \pi \alpha_{\rmii{QED}} Z^2 F^2_{\rmii{SI}}(E_R) \left\lbrace a_\gamma^2 \frac{8 m_A}{E_R}    \right. 
   \nonumber
   \\
   & \left. +  b_\gamma^2 \left[ 4 v^2 \left( \frac{m_A}{m_S}\right)^2 - 2 \left( 1+ 2 \frac{m_A}{m_S}\right)  \frac{m_A( E_R + 2 \DmVS)}{m_S^2}\right]
    \right. 
   \nonumber
   \\
   & +
   \left. a_\gamma b_\gamma \frac{8 m_A}{E_R} \left( \frac{m_A E_R}{m_S^2} - \frac{\DmVS}{m_S} \right) \right\rbrace  \, ,
   \label{nuc_matrix_element_NR}
\end{align}
where $Z$ denotes the nuclear charge, and $F_{\rm SI}(q^2)$  is the nuclear form factor. Due to the very mild running over the relevant momentum-transfer range, we use the fixed value $\alpha_{\rmii{QED}}=1/137$ in the following.  We adopt the Helm parametrisation \cite{Lewin:1995rx} for the spin-independent Xenon form factor when presenting our results for the differential recoil rate. 
We have checked that there are no major differences upon using the more recent form factor given in ref.~\cite{Vietze:2014vsa}.

The expression above makes the contributions of the two operators and their interference explicit. Most importantly, the term proportional to $a_\gamma^2$ exhibits a $1/E_R$ enhancement that is not suppressed by any small ratio of masses or other small scales. By contrast, the dimension-five operator $S V_{\mu} \partial_\nu F^{\mu \nu}$, whose contribution is proportional to $b_\gamma$, induces a contact-like interaction without a $1/E_R$ enhancement. 
Although the Wilson coefficient $b_\gamma$ is somewhat larger than $a_\gamma$, this difference is insufficient to compensate for the strong hierarchy between the contributions associated with the two operator structures.\footnote{The hierarchy between the squared matrix-element contributions can be illustrated by setting $a_\gamma=b_\gamma=1$. For the parameter space of interest in our analysis, we find $|\mathcal{M}_b|^2/|\mathcal{M}_a|^2 \approx 10^{-9}$ and $\mathrm{Re}(\mathcal{M}_a\mathcal{M}_b^\dagger)/|\mathcal{M}_a|^2 \approx 10^{-4}$.}
Therefore, the phenomenological analysis  can be performed using only the leading term in the first line of eq.~\eqref{nuc_matrix_element_NR}.

The next step towards the experimental observable is to compute the differential recoil rate.  
Accounting for the minimum velocity required to produce a given recoil energy $E_R$, cf.~eq.~\eqref{v_min_eq}, and using the differential scattering cross section in eq.~\eqref{inelastic_DM_N}, with the nucleon replaced by a nucleus, the differential recoil rate per unit detector mass is
\begin{eqnarray}
          \frac{dR}{d E_R} = \frac{\rho_{\rmii{DM}}}{32 \pi \, m_S^3 m_A^2} \int_{v > v_{\rmii{min}}(E_R)} d^3 v \frac{f_{\rmi{lab}}(\bm{v})}{v} |\mathcal{M}_{\scriptscriptstyle SA \to VA}(E_R)|^2 \,  
          \label{NUdiff_rate}
    \end{eqnarray}
where $\rho_{\rmii{DM}}$ is the local dark matter density. We adopt $\rho_{\rmii{DM}} = 0.3 \; \text{GeV}/\text{cm}^3$ as our default choice in the following. The convolution over the incoming dark matter velocity is performed in the laboratory frame. The laboratory-frame distribution is related to the galactic frame by
\begin{align}
    f_{\rmi{lab}}(\bm{v}) = f_{\rmi{gal}}(|\bm{v}+\bm{v}_{\rmii{E}}|) \, ,
\end{align}
where $\bm{v}_{\rmii{E}}$ is the velocity of the Earth through the halo, and we take the time average speed $\vE = 267.8$ km/s, as 
suggested in \cite{Baxter:2021pqo} \footnote{In particular for larger mass splitting the time variation of the Earth's velocity matters and induces a notable annual modulation of the direct detection rate \cite{McCabe:2026crm}. }. We model the galactic-frame velocity distribution with the Standard Halo Model which reads
\begin{align}
    f_{\rm gal}(\mathbf{u})
=
\frac{1}{N_{\rm esc}\,\pi^{3/2}v_0^3}
e^{-u^2/v_0^2}
\Theta(v_{\rm esc}-u) \, ,
\end{align}
with the remaining velocity parameters being $v_0 = 238$ km/s and $v_{\rmi{esc}}= 544$ km/s \cite{Baxter:2021pqo}.  The normalisation factor $N_{\rmi{esc}}$ is given by
\begin{align}
    N_{\rmi{esc}}= \textrm{erf} \left( \frac{v_{\rmi{esc}}}{v_0}\right) - \frac{2}{\sqrt{\pi}} \frac{v_{\rmi{esc}}}{v_0} \textrm{exp} \left( -\frac{v^2_{\rmi{esc}}}{v^2_0} \right) \, .
\end{align}
Retaining the dominant contribution to the squared matrix element in eq.~\eqref{nuc_matrix_element_NR}, the differential recoil rate can be approximated as
\begin{align}
    \frac{d R}{d E_R} \simeq \alpha_{\rmii{QED}} Z^2 F_{\rmii{SI}}(E_R)^2 a_\gamma^2 \frac{\rho_{\rmii{DM}}}{m_A m_S^3  E_R} \int_{v > v_{\rmii{min}}(E_R)} d^3 v \frac{ f_{\rmi{lab}}(\bm{v}) }{v} \, .
    \label{diff_rate_approx}
\end{align}
With the Standard Halo Model ansatz for the velocity distribution, the velocity integral can be preformed analytically, see Appendix~\ref{sec:App_A} for details. Note, that the traditional result for the velocity integral reported in e.g. \cite{Lewin:1995rx} has to be corrected if $v_{\rmi{min}}>v_{\rmi{esc}}-\vE$.

\begin{figure}
    \centering
    \includegraphics[width=0.48\linewidth]{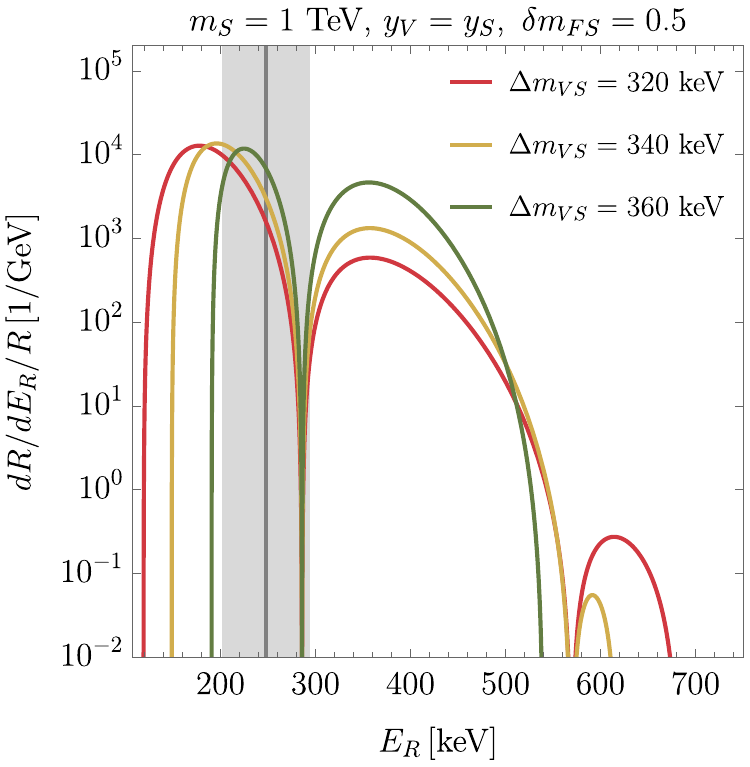}
    \includegraphics[width=0.48\linewidth]{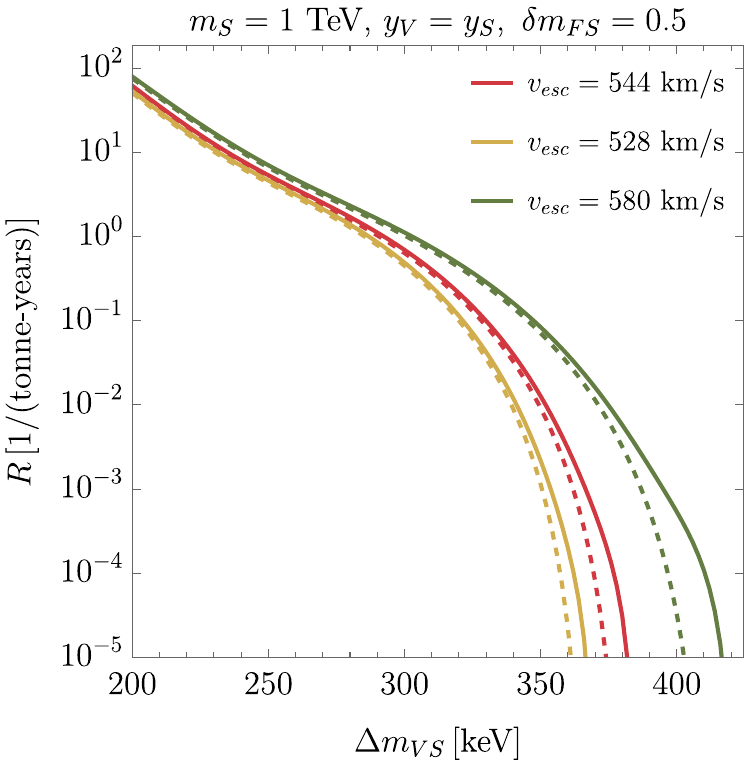}
    \caption{ Differential (Total) nuclear recoil rate for $m_S=1$ TeV, $\dmFS=0.5$ and equal Yukawa couplings, $\yS=\yV$, fixed by the observed relic density. We consider the light-lepton scenario, using the matching coefficient $a_\gamma^{(e,\mu)}$ and the Standard Halo Model parameters as described in the text.
    (Left) Normalized differential nuclear recoil rate as a function of the recoil energy for $\DmVS=320$, $340$, and $360$ keV. The shaded region indicates the recoil-energy interval of interest around the observed LZ event, $E^{\rmii{LZ}}_R = 248 \pm 46$~keV. (Right) Total rate as a function of $\DmVS$ for $v_{\rm esc}=528$, $544$  and $580$ km/s. The dashed lines include the effect of the experimental efficiency while solid lines correspond to the full theoretical prediction.}
    \label{fig:differential_R}
\end{figure}

In figure~\ref{fig:differential_R}, we show the differential recoil rate for benchmark values of the model parameters, considering the light-lepton option and hence the matching coefficient $a_\gamma^{(e,\mu)}$. We take equal Yukawa couplings, $\yS=\yV$, which are fixed by requiring the observed relic density, and set the dark matter mass to $m_S=1$ TeV in both panels. In the left panel, we 
show the differential recoil rate normalized to the total rate for three different values of the scalar--vector mass splitting, $\DmVS$. 
As expected, increasing $\DmVS$ restricts the range of allowed recoil energies and shifts the lowest allowed recoil energy $E_R$ upwards. A universal feature of the differential spectra is the minimum slightly below $E_R=300$ keV, which is imposed by the second diffraction minimum of the nuclear form factor. This leads to a double-peak structure, with one maximum of the recoil rate in the region consistent with the LZ observation and a second maximum at higher energies. Note that, due to the $1/E_R$ dependence of the squared matrix element, this second maximum is naturally suppressed in our model, such that we expect most events to arise in the preferred region. This helps alleviate the tension with the non-observation of events in the high-energy sideband, which has been pointed out as a challenge for the Higgsino interpretation of the LZ event~\cite{Rodd:2026tyn}. The right panel shows the dependence of the total rate on the splitting parameter $\DmVS$. As can be seen, $R$ is a steeply falling function of $\DmVS$ and varies by orders of magnitude. Interestingly, for the parameters considered here, a rate $R\sim 1/2.8/\mbox{tonne}/\mbox{year}$, corresponding to approximately one expected event in LZ when detector effects are ignored for simplicity, can be obtained for $\DmVS=310$ keV.

Before moving on, a word of caution regarding the Standard Halo Model is in order. This ansatz should be understood as a common baseline for the interpretation of direct detection data, rather than as an accurate description of the galactic velocity distribution on all scales. Inferring the distribution of dark matter within our galaxy is challenging, and the halo model is subject to significant uncertainties. Broadly speaking, these can be divided into two classes.

First, there are uncertainties related to the input parameters of the Standard Halo Model, such as $v_0$, $v_{\rm esc}$ and $\rho_{\rmii{DM}}$. A comprehensive discussion of different determinations of these parameters is given in \cite{Baxter:2021pqo}. The impact of these uncertainties on the interpretation of conventional searches for elastic spin-independent scattering is typically modest for heavy dark matter; see e.g. \cite{McCabe:2010zh}. However, the rate of interactions with large momentum transfer, such as that required to explain the LZ event, is more sensitive to these uncertainties, since only DM particles with high velocities contribute. To illustrate the impact on the recoil rate, we vary the escape velocity, which is the parameter with the largest uncertainty, between $580\, \mbox{km}/\mbox{s}$, the central value of \cite{Monari_2018}, and $528\, \mbox{km}/\mbox{s}$, the central value of \cite{Deason_2019}, as shown in the right panel of Fig.~\ref{fig:differential_R}. It is immediately apparent that the total recoil rate depends strongly on $v_{\rm esc}$ for mass splittings in the range preferred by the LZ event, and can vary by more than one order of magnitude.

Second, both observations and simulations indicate that the galactic halo is not fully virialized and contains distinct DM populations that cannot be described by a simple truncated Boltzmann distribution. For example, the kinematics of stars measured by Gaia indicate the presence of a distinct stellar population that likely originated from a merger with a sizable dwarf galaxy~\cite{Belokurov_2018}, commonly known as the ``Gaia sausage''. This merger is also expected to have left an associated DM population. The stars of the Gaia sausage can be used as tracers of the DM velocity distribution; see e.g. \cite{OHare:2019qxc} for a discussion of the general impact on direct detection signals and \cite{Zhang:2026hmg} for a more recent study focusing on the implications for LZ.

Another possibility considered in the literature is a distinct high-velocity DM component originating from the interaction between the Milky Way and the Large Magellanic Cloud. The size and shape of this component have been estimated using N-body simulations of Milky Way analogues. This contribution tends to enhance the high-velocity tail and can dominate at high velocities; see e.g. \cite{Smith-Orlik:2023kyl} for a general discussion and \cite{Bozorgnia:2026emp} for an application to LZ. As shown in \cite{Zhang:2026hmg} and \cite{Bozorgnia:2026emp}, allowing for such more general modifications of the halo model can significantly affect the interpretation of the rate in terms of particle-physics parameters. All things considered, we caution that any interpretation of the LZ event rate should be taken with a grain of salt and regarded as an order-of-magnitude estimate.

\subsection{Event rate}
\label{sec:event_rate}

In this section, we integrate the differential recoil rate and establish a connection with the LZ experimental results. As a first step, we determine the kinematically allowed recoil-energy range. For a given incoming dark matter speed $v$, the minimum and maximum recoil energies follow from the kinematic condition in eq.~\eqref{kin_mat_element}. In our numerical analysis, we evaluate these boundaries at the maximum laboratory-frame speed, $v_+=v_{\rm esc}+v_E$, yielding the full kinematically accessible recoil-energy window for the adopted halo model:
\begin{align}
E_R^{\min}(v_+)
&=
\frac{\mu_{SA}^2}{2m_A}
\left[
v_+-\sqrt{v_+^2-\frac{2\DmVS}{\mu_{SA}}}
\right]^2 \, , \quad
E_R^{\max}(v_+)
&=
\frac{\mu_{SA}^2}{2m_A}
\left[
v_++\sqrt{v_+^2-\frac{2\DmVS}{\mu_{SA}}}
\right]^2 \, .
\end{align}
The requirement that the square root be real sets an upper bound on the mass splitting for which the inelastic process is kinematically allowed,
\begin{align}
    \DmVS \leq \DmVS^{\rmi{max}} \equiv \frac{\mu_{\scriptscriptstyle SA}v_+^2}{2} \, .
    \label{eq:max_DmVS}
\end{align}
At this limiting value, the minimum and maximum recoil energies coincide, corresponding to scattering exactly at threshold. For larger mass splittings, the available kinetic energy of dark matter particles in the halo is insufficient to produce the heavier state \(V\).

The recoil-energy window is further restricted by the detector efficiency, for which we adopt the function given in ref.~\cite{LZ:2026axp}, denoted by $\epsilon_{\rmii{LZ}}(E_R)$. It exhibits a flat region, with $\epsilon_{\rmii{LZ}}(E_R) \simeq 0.95$ for $15 \leq E_R \leq 245$ keV, and a sharp cutoff at $E_R \geq 300$ keV. The total recoil rate  as seen by the detectors is then obtained by integrating the differential rate over this energy window
\begin{align}
    R = \alpha_{\rmii{QED}} Z^2 \frac{\rho_{\rmii{DM}} a_\gamma^2}{m_A m_S^3}
    \int_{E_R^{\min}(v_+)}^{E_R^{\max}(v_+)}
    dE_R\, \frac{\epsilon_{\rmii{LZ}}(E_R)F_{\rmii{SI}}(E_R)^2}{E_R}
    \int_{v > v_{\min}(E_R)} d^3v\,
    \frac{f_{\rmii{lab}}(\bm{v})}{v} \, .
\end{align}
Upon including the exposure $\mathcal{E}=2.84$ tonne-year used in the experimental analysis \cite{LZ:2026axp}, we obtain the predicted number of events as a function of the model parameters 

\begin{figure}
    \centering
    \includegraphics[width=0.34\linewidth]{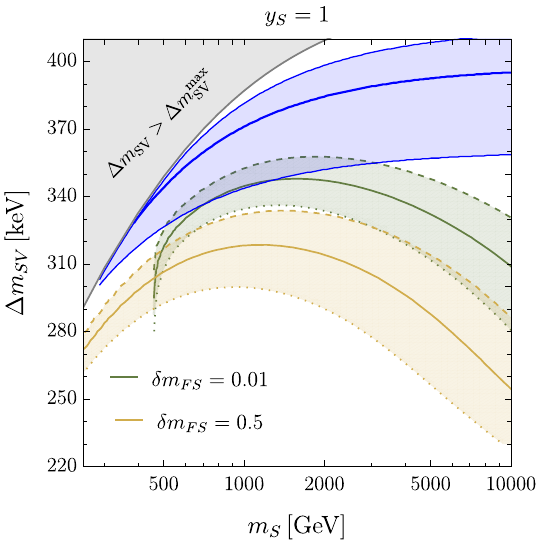}
     \includegraphics[width=0.315\linewidth]{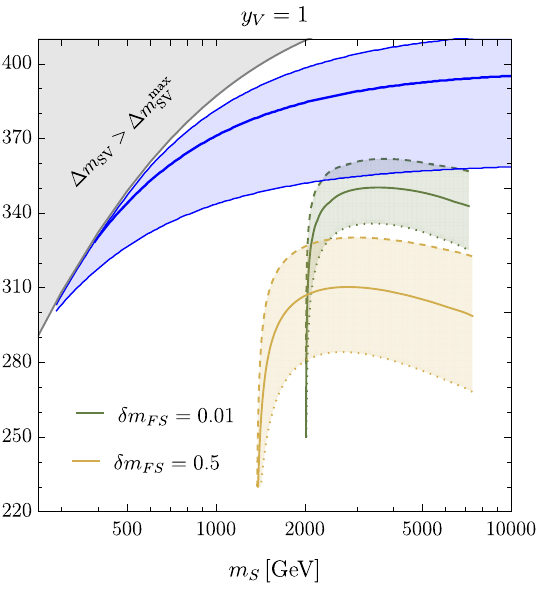}
      \includegraphics[width=0.315\linewidth]{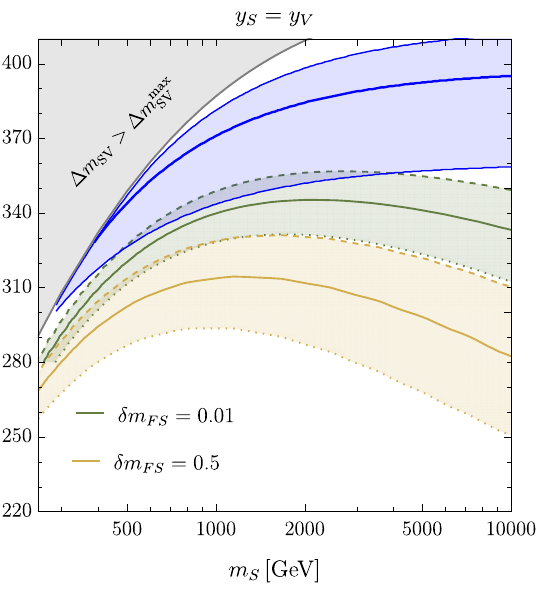}
    \caption{Contours of the predicted number of events for the three coupling assignments. The solid ochre and olive lines correspond to $N_{\rmii{events}}=1$, while the dashed and dotted lines correspond to $N_{\rmi{events}}=1/3$ and $N_{\rmi{events}}=3$, respectively. The blue shaded region, bounded by thick solid blue lines, indicates the parameter space where the maximum of the predicted differential recoil rate lies within the LZ recoil-energy interval, $E_R^{\rmii{LZ}}=248\pm46$ keV. Two values of the scalar–fermion mass splitting are considered: $\dmFS=0.5$ (ochre) and $\dmFS=0.01$ (green). The grey shaded region is kinematically forbidden.}
    \label{fig:paramater_mod_nf1}
\end{figure}

\begin{figure}
    \centering
    \includegraphics[width=0.34\linewidth]{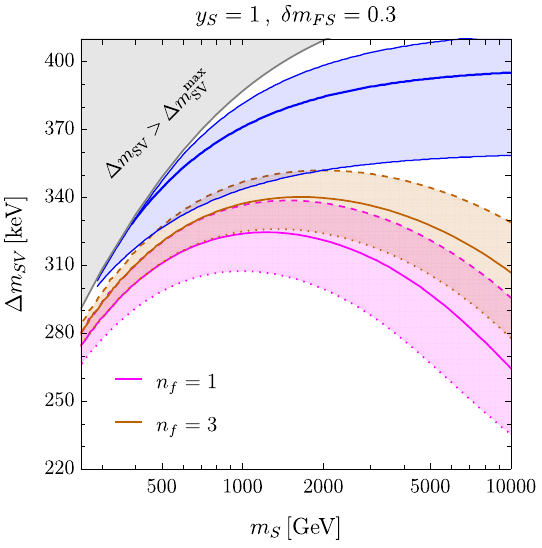}
     \includegraphics[width=0.315\linewidth]{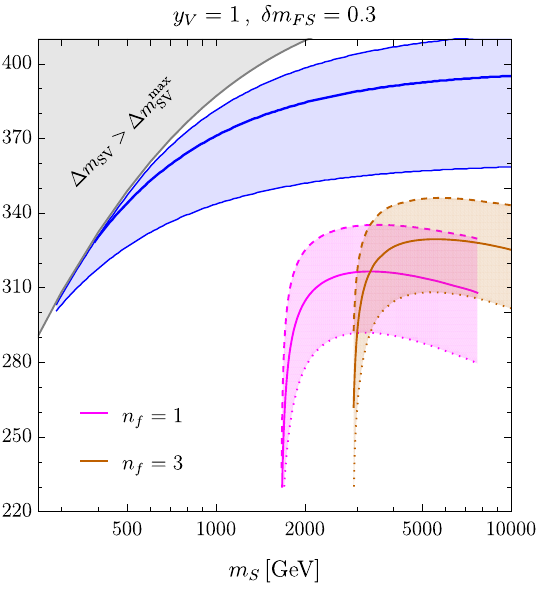}
      \includegraphics[width=0.315\linewidth]{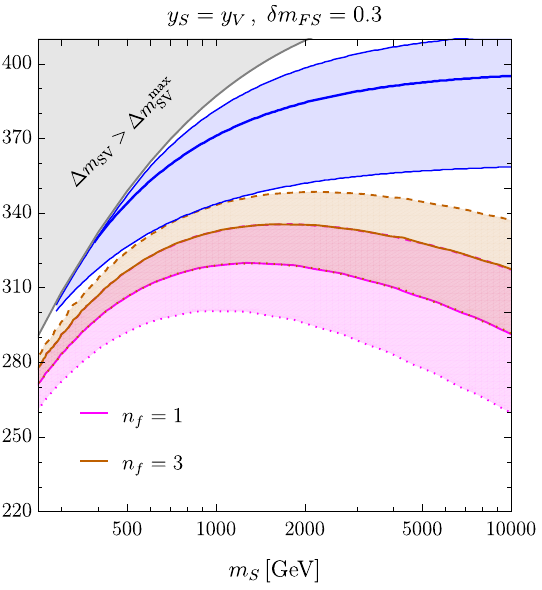}
    \caption{Contours of the predicted number of events for the three coupling assignments. The blue and grey shaded regions are as in Fig.~\ref{fig:paramater_mod_nf1}. The scalar–fermion mass splitting is fixed to $\dmFS=0.3$. The magenta and brown lines correspond to the dark bosons coupling to a single SM lepton flavour ($n_f=1$) or to all three lepton flavours with equal-strength diagonal couplings ($n_f=3$), respectively.}
    \label{fig:paramater_mod_nf3}
\end{figure}

Figures~\ref{fig:paramater_mod_nf1} and \ref{fig:paramater_mod_nf3} summarize the interplay between the leptophilic DM model and the high-energy recoil event reported by the LZ collaboration. The solid gray line indicates the maximum mass splitting that allows for inelastic scattering for the Standard Halo Model with our default parameters, cf.~eq.~\eqref{eq:max_DmVS}. No events can be generated in the corresponding gray-shaded region, where $\DmVS > \DmVS^{\rmi{max}}$. The solid blue line identifies the combinations of DM mass and vector--scalar mass splitting for which the differential nuclear recoil rate is maximal at the experimentally observed recoil energy. The blue band accounts for the uncertainty in the measured recoil energy and displays the interval in which the maximum lies within $E^{\rmii{LZ}}_R=248\pm46$ keV. The blue region therefore identifies the portion of the $(m_S,\DmVS)$ parameter space where the recoil spectrum is most compatible with the LZ event.

Each figure contains three panels, corresponding to the representative coupling assignments considered in our study: $\yS=1$, $\yV=1$, and $\yS=\yV$, from left to right. 
In all panels the value of the free coupling has been fixed by requiring that the relic density from  thermal freeze-out is consistent with the Planck measurement of $\Omega_{\rmii{DM}}h^2=0.12(12)$.
We first discuss the contours in Fig.~\ref{fig:paramater_mod_nf1}, where we consider two DM--mediator mass splittings, $\dmFS=0.5$ (ochre) and $\dmFS=0.01$ (green). The solid contours indicate the parameter space yielding one expected event, while the dashed and dotted contours correspond to $N_{\rmi{events}}=1/3$ and $N_{\rmi{events}}=3$, respectively. The comparison illustrates how a smaller mediator mass splitting leads to a larger event rate, and vice versa.
The relative position of the green and ochre contours is mainly determined by the dependence of the operator coefficients $a_\gamma$ on the mass splitting. In particular, the approximate scaling $a_\gamma\propto 1/(1+\dmFS)$ enhances the rate for small splittings, while suppressing it when $\dmFS$ is of order unity. Conversely, the relic-density constraint requires smaller Yukawa couplings when coannihilations with the fermion mediator are included. However, this effect is subleading.
For the small $\delta m_{FS}$ benchmark, the predicted recoil spectrum shows reasonable agreement with the LZ event in both its shape and expected rate, while the required couplings remain consistent with a thermal relic abundance.

It is worth noting the rather different shape of the contours in the case $\yV=1$. This is caused by two main effects. First, there is a lower bound on the dark matter mass from thermal freeze-out, 
since the processes driving the annihilations are already quite efficient for $\yV=1$. Consequently, small DM masses would lead to an excessively large annihilation cross section. On the other hand, as we move towards larger DM masses, keeping $\yV$ fixed to unity requires increasingly large values of $\yS$. These can contribute efficiently only through the coannihilation process $SV\to\ell\bar{\ell}$, since scalar-pair annihilation is $d$-wave suppressed. As a result, one rather quickly reaches the perturbative unitarity bound $\yS=\sqrt{4\pi}$, which is where we terminate the band. One may track the upper-mass end points for the $\dmFS=0.5$ and $\dmFS=0.01$ respectively in figure~\ref{fig:relic_density}, left panel, and notice that the largest mass  with a perturbative coupling is very similar for these two mass splittings.

Let us now turn to Fig.~\ref{fig:paramater_mod_nf3}, which 
focuses on the dependence on the number of lepton flavours and on the coupling assignments. The three panels correspond to the same representative choices of couplings as before, namely $\yV=1$, $\yS=\yV$, and $\yS=1$. In all panels, we fix the DM--mediator mass splitting to $\dmFS=0.3$, and display the case for $n_f=1$ and $n_f=3$ with magenta and brown curves respectively. 
Allowing DM to couple to all SM right-handed leptons leads to a coherent enhancement of the loop-induced matching coefficient, $a_\gamma\to n_f a_\gamma$, where $n_f$ denotes the number of lepton flavours. Hence, the rate acquires a $n_f^2$ enhancement with respect to the single-flavor case.  For this choice of the splitting, $\dmFS=0.3$, we are away from the coannihilation strip involving the fermion mediator, and including the coupling to all flavours amounts to modifying the cross sections in eqs.\eqref{SS_ll}-\eqref{VV_ll} by an overall factor of $n_f$. Taking this into account, one again finds that smaller Yukawa couplings are required to reproduce the observed relic density. However, as in the previous case, this effect is subdominant to the enhancement arising from the coherent flavour sum at the level of the squared matrix element. This boost of the scattering rate is sufficient to establish marginal consistency between the thermal history, the rate and the signal shape for two of the coupling assignments whereas the rate falls slightly short in the $y_V=1$ case.

Overall, our results demonstrate that a differential recoil rate consistent with the LZ signal can be explained by thermal dark matter in this model in various parts of the parameter space.

\section{Other constraints}
\label{sec:other_constraints}
Having established that the model under consideration is able to explain the tentative LZ signal and the thermal relic density simultaneously, it is important to investigate whether the preferred parameter space is consistent with other experimental searches. In the following, we consider indirect searches for gamma-ray lines and high energy neutrino emission from the Sun as well as collider searches at the LHC.

\subsection{Indirect detection}
\label{sec:indirect_det}

\subsubsection*{Gamma-ray lines}
Dark scalars can annihilate into photons through radiative processes, which fall into two classes: the $2\to3$ annihilation channel $SS\to \ell \bar{\ell} \gamma$, corresponding to internal bremsstrahlung, and the loop-induced $2\to2$ channels $SS\to\gamma\gamma$ and $SS\to Z\gamma$. The corresponding cross sections have been computed in ref.~\cite{Ibarra:2014qma}, and we adopt their results, with the appropriate modification to account for the lepton-flavour multiplicity.

Most notably, these radiative processes contain velocity-independent contributions and can therefore remain relevant in the present-day Universe, despite the additional suppression by $\alpha_{\rmii{QED}}$. The relevant cross sections scale as $\yS^4 \alpha_{\rmii{QED}}/m_S^2$ and $\yS^4 \alpha_{\rmii{QED}}^2/m_S^2$ for internal bremsstrahlung and loop-induced processes, respectively. Despite the additional electromagnetic suppression, the loop-induced channels can become competitive for sufficiently large mediator–DM mass ratios, owing to the different dependence of the cross sections on the mediator mass.
The resulting photon spectra are hard and can therefore be effectively searched for with gamma-ray telescopes. We compare the predicted cross sections with gamma-ray constraints from Fermi-LAT~\cite{Fermi-LAT:2015kyq} and H.E.S.S.~\cite{HESS:2018cbt,HESS:2026ila}, distinguishing between the two experiments and the corresponding spectral signatures.

To this end, we evaluate the cross sections for the processes with photons in the final state over the cosmologically viable parameter space. We consider the dark matter energy density as a function of four parameters, $\Omega_{\rmii{DM}}h^2(\yS,\yV,m_S,\dmFS)$, and use the observed relic density to fix the dark scalar mass as a function of the scalar Yukawa coupling, for fixed mediator mass splittings and the three coupling assignments discussed in sec.~\ref{sec:freeze_out}. 

We now assess the gamma-ray constraints on the model using the results from Fermi-LAT~\cite{Fermi-LAT:2015kyq} and H.E.S.S.~\cite{HESS:2018cbt,HESS:2026ila}. The Fermi-LAT limits cover only part of the parameter space considered here, in the mass interval $100{\rm GeV}\lesssim m_S\lesssim294~{\rm GeV}$. For this range, we consider the total radiative annihilation rate, obtained by summing the cross sections for $SS\to \ell \bar\ell\gamma$, $SS\to\gamma\gamma$ and $SS\to Z\gamma$, and compare it with the corresponding Fermi-LAT limits. Since these channels produce different photon spectra, this procedure should not be interpreted as a dedicated spectral analysis or as a statistically consistent combination of the individual constraints. Instead, it gives an approximate indication of the sensitivity of Fermi-LAT to the overall radiative annihilation rate.

At higher masses, $m_S\gtrsim310~{\rm GeV}$, we consider the latest H.E.S.S. results, which constrain monochromatic gamma-ray lines over a mass range extending into the multi-TeV regime~\cite{HESS:2026ila}. In this case, the relevant contributions come from the two line-producing channels, $SS\to\gamma\gamma$ and $SS\to Z\gamma$. We therefore compare the predicted line cross section,
\begin{align}
\langle \sigma \vrel \rangle_{\rm line}
=\langle\sigma \vrel \rangle_{\gamma\gamma}
+\frac{1}{2}\langle\sigma \vrel \rangle_{Z\gamma},
\end{align}
with the H.E.S.S. upper limit evaluated at the DM mass $m_S$. The factor of $1/2$ accounts for the fact that the two final states produce different numbers of monochromatic photons. Thus, in contrast to the approximate comparison adopted for Fermi-LAT, the H.E.S.S. constraint can be applied directly to the line cross section defined above.

\begin{figure}
    \centering
    \includegraphics[width=0.5\linewidth]{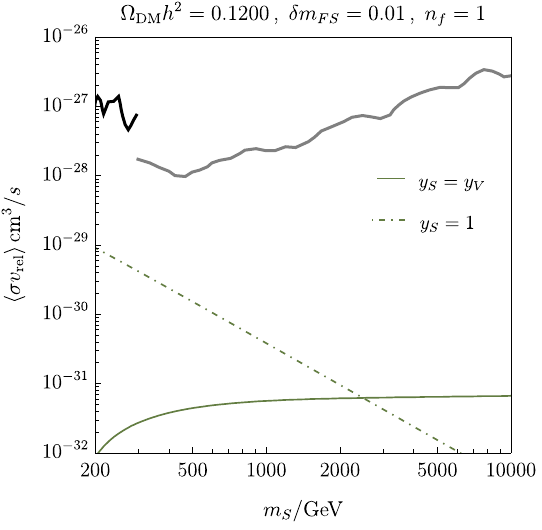}\includegraphics[width=0.5\linewidth]{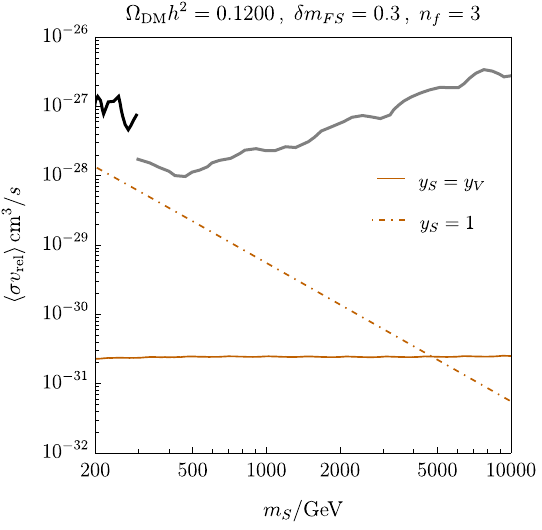}
    \caption{Predicted cross sections for annihilation channels with one or two photons in the final state, compared with the current constraints from Fermi-LAT (black-solid line) and H.E.S.S (gray-solid line). The left and right panels show the results for $\dmFS=0.01$, $n_f=1$ and $\dmFS=0.3$, $n_f=3$, respectively. The coupling assignment is indicated in each panel. In both cases, the parameter space favoured by the relic-density requirement and the LZ high-energy recoil is not excluded by current indirect-detection searches.}
    \label{fig:ID_gamma_ray}
\end{figure}

Our findings can be summarised as follows. Overall, the constraints on the multi-component leptophilic model are rather mild. In Fig.~\ref{fig:ID_gamma_ray}, we show the predicted cross sections for final states containing one or two photons, and compare them with the limits from H.E.S.S. and Fermi-LAT.  We consider the two mass splittings and coupling assignments that provide the best agreement with the LZ direct-detection signal, namely $\dmFS=0.01$ with $n_f=1$ and $\dmFS=0.3$ with $n_f=3$. The corresponding results are shown in the left and right panels, respectively. As can be seen, current indirect-detection sensitivities are insufficient to probe the parameter space favoured by the relic-density requirement and the high-energy nuclear recoil.

It is also interesting to compare our results with the recent phenomenological analysis of the leptophilic model containing only the scalar and fermionic components~\cite{Biondini:2026dpm}, in which gamma-ray constraints were found to be more relevant. The main difference can be traced back to the Yukawa couplings required to reproduce the observed relic density. In the present multi-component scenario, the vector-assisted freeze-out mechanism allows the observed relic abundance to be obtained with smaller values of the scalar Yukawa coupling, $\yS$. Since the indirect-detection cross sections depend on $\yS$, this leads to a suppression of the corresponding gamma-ray signals and, consequently, to weaker constraints from gamma-ray searches.

\subsubsection*{Solar capture and neutrinos from the Sun}
Inelastic scattering of dark matter on nuclei can lead to the capture of dark matter particles in the Sun~\cite{Nussinov:2009ft,Menon:2009qj}. Over time, the captured dark matter population accumulates, creating a significant local overdensity within the Sun. This can considerably enhance the annihilation rate, in particular if subsequent scattering allows the dark matter distribution to contract to a radius smaller than that associated with the initial capture process. If these processes are efficient, the dark matter capture rate can reach equilibrium with the annihilation rate, such that limits on high-energy neutrino production in the Sun can be used to constrain the direct detection rate directly.

For dark matter masses above $\mathcal{O}(10)$ GeV, the most stringent limits for the elastic scattering case come from IceCube~\cite{IceCube:2025fcu}. Reinterpreting these results for inelastic scattering has been found to be highly constraining for certain explanations of the LZ events~\cite{Pospelov:2026ewn,Nguyen:2026lui,DiMauro:2026}, most notably the thermal Higgsino model. In our case, however, this is not a concern. First, for dark matter coupled to electrons or muons, the flux of high-energy neutrinos produced in the annihilations is highly suppressed. While the electron channel does not produce any neutrinos at leading order, neutrinos can in principle arise from muon decays. These are, however, very soft, as muons rapidly lose energy in the solar medium and retain only a negligible fraction of their original kinetic energy before decaying; see e.g.~\cite{Baratella:2013fya}. Thus, solar capture does not provide relevant constraints on dark matter candidates coupled exclusively to these lepton flavours.

The situation is somewhat more complicated for couplings to taus, which decay before losing a significant amount of energy. Consequently, annihilations into the $\tau^+\tau^-$ channel lead to a hard neutrino flux. Nevertheless, even in this case, it is questionable whether solar capture provides a meaningful constraint, given that the annihilation cross section for the real scalar dark matter candidate considered here is $d$-wave suppressed. This considerably increases the time required to reach equilibrium between capture and annihilation, thereby suppressing the resulting neutrino flux. Given the current uncertainties, we consider the unconstrained light-lepton-flavour variants of our model sufficient to demonstrate its viability as an explanation of the LZ signal. A more detailed study of solar capture in the tau case might be warranted if the signal is further substantiated.

\subsection{Collider}
The charged fermionic mediator can be pair-produced at the LHC through Drell--Yan processes, leading to final states with two opposite-sign leptons and missing transverse momentum. This signature has been studied in leptophilic dark matter models with scalar or vector dark matter and a vector-like fermion mediator \cite{Barducci:2018esg}. Existing ATLAS and CMS searches constrain mediator masses up to approximately 400 GeV for electron- and muon-philic scenarios, while the corresponding reach is weaker for tau-philic interactions.
 Irrespective of the lepton flavor, the sensitivity deteriorates when the mediator and dark matter masses are close, due to the soft leptons produced in mediator decays.
More recently, a comprehensive assessment of collider searches for $t$-channel dark matter models has highlighted the sensitivity of Run-2 searches to this topology, as well as the challenges associated with compressed mass spectra \cite{Arina:2025zpi}.  Given the mass range and mediator mass splittings considered in our analysis, collider searches are not expected to provide significant additional constraints on most of the parameter space of interest with $m_S \gtrsim 300$ GeV. 
\section{Summary and Conclusions}
\label{sec:conclusions}

In this work, we have investigated the direct-detection phenomenology and cosmological implications of a leptophilic multicomponent dark sector, motivated by the high-energy nuclear-recoil event reported by the LZ experiment. Our model belongs to the class of $t$-channel dark matter scenarios, in which dark states interact with SM leptons through a charged fermion mediator. These frameworks have been extensively explored in collider, direct-detection and indirect-detection studies, typically considering a single dark matter candidate. Here, we extended this setup by combining a real scalar and a real vector, coupled to right-handed SM leptons through a common vector-like fermion mediator. This provides a simple realization of multicomponent dark matter in which transitions between dark states can give rise to the high-energy recoil event observed by LZ.

As the only SM particles to which the dark matter states couple directly are leptons, the leading contributions to both elastic and inelastic dark-matter–nucleus scattering arise at higher order in perturbation theory. The dominant contributions are generated by loop-induced couplings of the dark-sector states to photons, which are most conveniently described using an EFT.
A central result of our analysis is that the inelastic scalar--vector transition is mediated by dimension-five operators, whereas the leading photon-mediated elastic interactions arise at dimension six. This hierarchy naturally enhances inelastic scattering relative to elastic scattering and allows the model to produce high-energy nuclear recoils while suppressing the low-energy signal constrained by conventional direct-detection searches. Moreover, the dominant transition operator predicts a differential recoil spectrum with a characteristic $1/E_R$ dependence, arising from photon exchange. This results in a softer recoil spectrum than in many other models, suppressing the scattering rate in LZ's high-energy sideband. The non-observation of events at higher energies is therefore also consistent with this interpretation of the LZ event.

We have connected this direct-detection signature to the thermal history of the dark sector and identified the parameter choices compatible with the observed relic density. We find that thermally produced dark matter can yield an event rate of the order required to account for the LZ observation. The same dimension-five operator also induces the decay of the heavier vector into the lighter scalar and a photon. The resulting lifetime is typically of order seconds, implying that the vector leaves the detector before decaying.
Unfortunately, these decays do not seem to lead to testable effects on Big Bang Nucleosynthesis.

We have examined the recoil spectrum and event rate in detail and find that the differential rate peaks at the energies preferred by LZ for mass splittings of $300$--$400$~keV, depending on the dark matter mass. Our results show that the shape and rate of the direct-detection signal, as well as the thermal relic density, can simultaneously match the observations in our model. However, we caution that the recoil rate is subject to significant astrophysical uncertainties, e.g. variations in the escape velocity or possible deviations of the velocity distribution from the simple Standard Halo Model ansatz. These can change the rate by a factor of 10 or more for the mass splittings considered here. As in many interpretations of the LZ event, our predicted event rate should therefore be regarded as an order-of-magnitude estimate rather than a precise prediction. Our results should thus be viewed as a proof of principle. We have also checked additional constraints and find that existing collider and indirect-detection constraints do not significantly restrict the parameter space relevant to this analysis.

LZ continues to collect data, while both PandaX and XENONnT are also expected to be sensitive to this experimental signature. We hope that future measurements will help clarify the nature of this intriguing observation by LZ.

\section*{Acknowledgments}
SB is grateful to Beatrice Lace for bringing the news of the LZ announcement to the dinner table on the very same day, and  Marc Schumann for helpful discussions. 
LT thanks the Institute of Physics of the University of Freiburg for its hospitality and the Erasmus Traineeship program. The work of CH is supported by the Deutsche Forschungsgemein-
schaft (DFG, German Research Foundation) - RTG 3186 - 556371633. 
\appendix
\renewcommand{\thesection}{\Alph{section}}
\renewcommand{\thesubsection}{\Alph{section}.\arabic{subsection}}
\renewcommand{\theequation}{\Alph{section}.\arabic{equation}}

%
\section{Velocity integrals}
\label{sec:App_A}

In this appendix, we provide the expression for the relevant velocity integral entering the calculation of the nuclear recoil rate in section~\ref{sec:direct_det}. 
The velocity distribution in the laboratory frame is 
\begin{align}
    f_{\rmi{lab}}(\bm{v}) = f_{\rmi{gal}}(|\bm{v}+\bm{v}_{\rmii{E}}|)
\end{align}
where $\bm{v}_{\rmii{E}}$ is the velocity of the Earth through the halo. Choosing the $z$-axis along $\bm{v}_{\rmii{E}}$, the condition $|\bm{v}+\bm{v}_{\rmii{E}}| < v_{\rmi{esc}}$ gives the following constraint on the cosine ($\cos{\theta} \equiv c_\theta$):
\begin{align}
    c_\theta < \frac{v^2_{\rmi{esc}}-v^2-\vE^2}{2v v_{\rmii{E}}} \equiv c_{\theta}^{\rmi{max}} .
\end{align}
Since $-1 \leq c_\theta \leq 1$, different possibilities arise depending on the value of $v$ and, consequently, on the corresponding value of $c_{\theta}^{\rmi{max}}$. Two cases can be distinguished:
\begin{itemize}
    \item[i)] $c_{\theta}^{\rmi{max}} \geq 1$ for $v \leq v_{\rmi{esc}}-\vE$. In this case, the angular integral covers the full range of $c_{\theta}$, i.e.~$-1 \leq c_\theta \leq 1$.
    \item[ii)] $-1 \leq c_\theta < c_\theta^{\rmi{max}}$ for $v_{\rmi{esc}}-\vE < v < v_{\rmi{esc}}+\vE$. In this case, the angular integral is restricted to the range $-1 \leq c_\theta < c_\theta^{\rmi{max}}$.
\end{itemize}

The velocity integral relevant for the nuclear recoil rate, which enters the differential recoil rate in eq.~\eqref{diff_rate_approx} and features a minimum velocity as its lower integration boundary, reads
\begin{align}
            &\int d^3v \frac{f_\text{lab}(\bm{v})}{v} \nonumber  \\
            &=\frac{2\pi}{N_\text{esc}\pi^{3/2}v_0^3}\int_{v_\text{min}}^{\infty} dv \, v \int_{-1}^{1} d c_\theta \exp\left(-\frac{v^2+v_E^2+2v \vE\cos{\theta}}{v_0^2}\right)\Theta(v_\text{esc}-v) \, .
        \end{align}
The integral can be evaluated analytically, and the result depends on the value of $v_{\rmi{min}}$.
For $v_\text{min}<v_\text{esc}-\vE$, we find
\begin{align}
     \int d^3v \frac{f_\text{lab}(\bm{v})}{v} &=  \frac{2\pi}{N_\text{esc}\pi^{3/2}v_0^3}\int_{v_\text{min}}^{v_\text{esc}-v_E} dv \, v \int_{-1}^{1} d c_\theta \exp\left(-\frac{v^2+\vE^2+2v \vE c_\theta}{v_0^2}\right)\nonumber
     \\
            &+\frac{2\pi}{N_\text{esc}\pi^{3/2}v_0^3}\int_{v_\text{esc}-v_E}^{v_\text{esc}+v_E} dv \, v \int_{-1}^{c_\theta^{\text{max}}} d c_\theta \exp\left(-\frac{v^2+v_E^2+2v \vE c_\theta}{v_0^2}\right)\nonumber
     \\
     &= \frac{1}{2 N_\text{esc} \, \vE v_0 \sqrt{\pi}} \left\lbrace - 4 \vE e^{-v^2_{\rmii{esc}}/v_0^2} + v_0 \sqrt{\pi} \left[ \text{erf} \left( \frac{v_{\rmi{min}}+\vE}{v_0} \right) - \text{erf} \left( \frac{v_{\rmi{min}}-\vE}{v_0} \right)  \right] \right\rbrace  \, , 
     \label{v_integral_case1}
\end{align}
whereas for $v_\text{min}>v_\text{esc}-v_E$, only the second angular integral appearing in eq.~\eqref{v_integral_case1} contributes, yielding
\begin{align}
     \int d^3v \frac{f_\text{lab}(\bm{v})}{v} &=  
            \frac{2\pi}{N_\text{esc}\pi^{3/2}v_0^3}\int_{v_\text{min}}^{v_\text{esc}+\vE} dv \, v \int_{-1}^{c_\theta^{\text{max}}} d c_\theta \exp\left(-\frac{v^2+\vE^2+2v \vE c_\theta}{v_0^2}\right)
            \nonumber\\
            &=\frac{1}{2 N_\text{esc} \, \vE v_0 \sqrt{\pi}}  \Bigg(-2 e^{-v^2_{\rmii{esc}}/v_0^2} (\vE + v_\text{esc} - v_\text{min})  \nonumber \\
            &\qquad+ \sqrt{\pi} v_0 \bigg[\text{erf}\left(\frac{v_\text{esc}}{v_0}\right) + \text{erf}\left(\frac{\vE - v_\text{min}}{v_0}\right)\bigg]\Bigg) \, .
              \label{v_integral_case2}
\end{align}
The result in eq.~\eqref{v_integral_case1} is equivalent to the canonical result of ref.~\cite{Lewin:1995rx}. The second case, cf.~\eqref{v_integral_case2}, was not included in that work, likely because such high-recoil-energy events were not of particular interest at the time.

\section{Coannihilations with the mediator}
\label{sec:App_B}

If the mass splitting between the dark bosons and the fermion is also small, the vector-like lepton will be present during freeze-out as well. This can be easily incorporated, since Eqs.~\eqref{BE_start} and \eqref{eq:sigma_eff_general} are equally valid in this case, provided that we understand the total dark-sector number density as the sum of the scalar, vector and fermionic number densities,
\[
n=n_{\scriptscriptstyle S}+n_{\scriptscriptstyle V}+2n_{\scriptscriptstyle F},
\]
where the factor of two for the mediator accounts for the particle and antiparticle.

The main modification concerns Eq.~\eqref{eff_Xsection_coann_real_S_v2}, where additional channels appear. For our purposes, we can take the mass splittings between the scalar and the vector to be equal and, in any case, much smaller than the mass splitting with the fermionic mediator. Hence, the effective cross section is modified as follows:
\begin{align}
\sigma_{\textrm{eff}} \vrel
&=
\frac{1}{\left[1+\frac{g_{\scriptscriptstyle V}}{g_{\scriptscriptstyle S}}+
2\frac{g_{\scriptscriptstyle F}}{g_{\scriptscriptstyle S}}
e^{-\DmFS/T}(1+\dmFS)^{\frac{3}{2}}\right]^2}
\left[
\sigma_{SS}\vrel
+2\frac{g_{\scriptscriptstyle V}}{g_{\scriptscriptstyle S}}\sigma_{SV}\vrel
+\frac{g_{\scriptscriptstyle V}^2}{g_{\scriptscriptstyle S}^2}\sigma_{VV}\vrel
\right.
\nonumber\\
&\quad\left.
+4\frac{g_{\scriptscriptstyle F}}{g_{\scriptscriptstyle S}}
e^{-\Delta m_{\scriptscriptstyle FS}/T}(1+\dmFS)^{\frac{3}{2}}
\sigma_{FS}\vrel
+4\frac{g_{\scriptscriptstyle F}g_{\scriptscriptstyle V}}{g^2_{\scriptscriptstyle S}}
e^{-\Delta m_{\scriptscriptstyle FS}/T}(1+\dmFS)^{\frac{3}{2}}
\sigma_{FV}\vrel
\right.
\nonumber\\
&\quad\left.
+2\frac{g_{\scriptscriptstyle F}^2}{g_{\scriptscriptstyle S}^2}
e^{-2\DmFS/T}(1+\dmFS)^3
\left(\sigma_{F\bar F}\vrel+\sigma_{FF}\vrel\right)
\right].
\label{eff_cross_with_F}
\end{align}
As can be seen, the additional annihilation processes displayed in the second and third lines of Eq.~\eqref{eff_cross_with_F} are effectively suppressed for large relative mass splittings $\dmFS$, thereby recovering Eq.~\eqref{eff_Xsection_coann_real_S_v2}, whereas they may contribute to and enhance the cross section for $\dmFS\lesssim0.1$. This allows us to extend the study of the parameter space to the coannihilation strip. 

Determining the relic density still requires the annihilation cross sections with one or two fermionic mediators in the initial state. We compute these cross sections and collect the results below. They extend the scenarios in which the dark sector consists of either the real scalar and the vector-like lepton, or the real vector and the mediator. The fermion-antifermion and fermion-fermion (and its complex conjugate) annihilation processes have recently been computed for the scenario with scalar dark matter only \cite{Biondini:2026dpm}.

The leading velocity independent contributions to Eq.~\eqref{eff_cross_with_F} are collected below, assuming vanishing SM particle masses. These formulas refer to the one-flavour scenario:
\begin{align}
\sigma_{FV\to\ell\gamma}\vrel
&=
\frac{\alpha_{\rmii{QED}}\yV^2(5m_F^2+4m_Fm_V+2m_V^2)}
{24m_F^2m_V(m_F+m_V)},
\label{eq:sigma_VF_gamma}
\\[4pt]
\sigma_{FV\to\ell Z}\vrel
&=
\frac{\alpha_{\rmii{QED}}\yV^2}
{96c_W^2s_W^2m_F^2m_V(m_F+m_V)}
\nonumber\\
&\quad\times
\left[
3c_W^4m_F^2
+2c_W^2s_W^2m_F(m_V-5m_F)
+s_W^4(7m_F^2+24m_Fm_V+8m_V^2)
\right].
\label{eq:sigma_VF}
\end{align}
For the fermion-antifermion annihilation channel, we obtain
\begin{align}
\sigma_{F\bar F\to\ell\bar\ell}\vrel
&=
\frac{5\pi \alpha_{\rmii{QED}}^2}{c_W^2m_F^2}
-\frac{\alpha_{\rmii{QED}}}{8c_W^2}
\left(
\frac{\yS^2}{m_F^2+m_S^2}
+\frac{2\yV^2}{m_F^2+m_V^2}
\right)
\nonumber\\
&\quad+
\frac{m_F^2}{32\pi}
\left[
\frac{4\yS^2\yV^2}{(m_S^2+m_V^2)(m_F^2+m_V^2)}
+\frac{\yS^4}{(m_F^2+m_S^2)^2}
+\frac{6\yV^4}{(m_F^2+m_V^2)^2}
\right].
\label{eq:sigma_FFbar}
\end{align}
For fermion-fermion annihilation, the cross section is
\begin{align}
\sigma_{FF\to\ell\ell}\vrel
&=
\frac{m_F^2}
{16\pi(m_F^2+m_V^2)^2(m_F^2+m_S^2)^2}
\nonumber\\
&\quad\times
\left[
14\yV^4(m_F^2+m_V^2)^2
-8\yV^2\yS^2(m_F^2+m_V^2)(m_F^2+m_S^2)
+\yS^4(m_F^2+m_V^2)^2
\right].
\label{eq:sigma_FF}
\end{align}
Here we have used $e^2=4\pi \alpha_{\rmii{QED}}$, $c_W=\cos\theta_W$ and $s_W=\sin\theta_W$. Expressions for $\sigma_{FS}\vrel$ and additional channels entering $\sigma_{F\bar F}$ can be found in \cite{Biondini:2026dpm}.

In the $F\bar F$ and $FF$ ($\bar F\bar F$) annihilation channels, two types of processes contribute: those driven by electroweak interactions, involving couplings of order $e\approx0.3$, and those driven by Yukawa couplings, which may be of the same order or much larger. For the charged fermion mediator, near-threshold effects can potentially affect the annihilation cross sections. They can enhance or reduce the fermion-antifermion and fermion-fermion annihilation rates in the attractive and repulsive channels, respectively, through Sommerfeld factors \cite{Hisano:2004ds,Iengo:2009ni}. Moreover, the attractive channel may support bound states \cite{Detmold:2014qqa,vonHarling:2014kha,Petraki:2015hla}. Since the relevant SM gauge interactions are electromagnetic and involve $Z$-boson exchange, these effects are rather mild. Furthermore, there is a partial cancellation between the attractive and repulsive channels, which makes the overall impact of near-threshold effects on the relic density only a few per cent, as recently discussed in ref.~\cite{Biondini:2026dpm}.

\bibliographystyle{JHEP}
\bibliography{paper.bib}
\end{document}